\documentclass[ reprint,2,amsmath,amssymb,aps]{revtex4-1}

\usepackage{graphicx}
\usepackage{dcolumn}
\usepackage{bm}
\usepackage{natbib}
\usepackage[table]{xcolor}
\usepackage{tabularx}
\usepackage{array}
\usepackage{colortbl}
\usepackage{appendix}
\usepackage{amsmath}

\begin{document}

\preprint{APS/123-QE}

\title{Fisher information analysis for quantum-enhanced parameter estimation in an electromagnetically-induced-transparency spectrum with single photons}
\author{Pin-Ju Tsai$^{1,*}$, Lun-Ping Yuan$^{2,3}$, and Ying-Cheng Chen$^{2,4}$}
\email{tpinju@ncu.edu.tw}
\affiliation{$^{1}$Department of Physics, National Central University, Taoyuan City 320317, Taiwan
}
\affiliation{$^{2}$Institute of Atomic and Molecular Sciences, Academia Sinica, Taipei 10617, Taiwan
}
\affiliation{$^{3}$Department of Physics, National Taiwan University, Taipei 10617, Taiwan
}
\affiliation{$^{4}$Center for Quantum Technology, HsinChu 30013, Taiwan
}
\date{\today}
\begin{abstract}
Electromagnetically-induced-transparency (EIT) spectroscopy has been used for the development of  sensitive sensors for quantum metrology applications. The sensitivity of a sensor is strongly dependent on the measurement precision of the EIT spectrum. In this work, we present a theoretical study of spectral lines hape measurement of three-level $\Lambda$-type EIT media based on Fisher information (FI) analysis. Using two kinds of probing sources, the single-photon Fock state and the coherent state, we calculate the FI in an EIT medium and quantify the quantum advantage and limitations of the single-photon probe. The analysis of the FI structure also provides a clear picture for classifying the spectral line shape into two different regimes, EIT and Aulter-Townes splitting (ATS). This work provides a systematic analysis of the single-photon EIT spectrum, which provides essential knowledge of quantum sensing based on EIT and deepens our understanding of the spectral characteristics of the $\Lambda$-type media.
\end{abstract}
\maketitle
\section{Introduction}
The electromagnetically-induced-transparency (EIT) \cite{fleischhauer2005electromagnetically} effect plays an important role in modern quantum science and technology. It enables the control \cite{sagona2020conditional}, storage \cite{lvovsky2009optical} and manipulation of \cite{tsai2020quantum} the incident quantum state of electromagnetic fields based on the quantum interference between light and matter \cite{fleischhauer2000dark}. One special feature of the EIT medium is the spectrum transparent window  at the two-photon resonance condition. The theoretical 100\% transmission of a probe signal at the window makes it possible for the EIT medium to sense the changes in the spectral characteristics occurring due to the environmental variation while inhibiting signal attenuation. This means that, the EIT medium can be regarded as a sensitive quantum sensor and has been widely used in magnetometers \cite{scully1992high, zhang2016high, sun2017cavity, PhysRevA.62.013808}, electric field meters \cite{meyer2021optimal,meyer2022simultaneous,mohapatra2008giant,sedlacek2012microwave} and a velocimeter \cite{kuan2016large}.

In EIT-based sensing, the shift in the transparent window of the absorption spectrum which occurs in reaction to the change of the physical parameters allows it to be measured. Therefore, the accuracy of an EIT sensor is determined by how precisely  the EIT spectrum can be measured. However, spectroscopy utilizing classical probe source of a laser (coherent state) is typically limited by the so-called shot-noise limit (SNL). The SNL comes from the nature of the Poisson distribution of the coherent states in the photon number basis. For a detection method based on photon counting measurement, the coherent state with a mean photon number of $\bar{n}$ has an intrinsic standard deviation of photon number statistics of $\sqrt{\bar{n}}$, which bounds the accuracy of the absorption estimation and thus the precision of spectral measurement. In absorption spectroscopy, sub-Poissonian light has been introduced as the probe source to overcome the SNL  \cite{adesso2009optimal,li2021quantum,whittaker2017absorption} to gain the quantum advantage. Sub-Poissonian light has a relatively concentrated photon-number distribution, which results in a sharper contrast in the photon-number distribution after absorption, compared to a classical light source. The quantum advantage for parameter estimation is even more significant in the weak absorption regime \cite{allen2020approaching,allen2019quantum}. 

In EIT-based sensing, most information is obtained around the transparent window, which has a relatively low absorption, on the probe. Thus, it is expected that using a sub-Poisson light (e.g. a single photon source) to conduct absorption spectroscopy will greatly increase the precision of frequency estimation of an EIT spectrum. In this paper, we study the use of a single-photon probe to boost the frequency estimation of the spectrum in a $\Lambda$-type EIT atomic system by analyzing the Fisher information (FI). We consider two different light sources in the single-photon Fock state (quantum case) and in the coherent state (classical case) as the probe to identify the quantum advantage and their limitations. Another interesting point, is that the FI structure with a single-photon probe can be used to classify the spectrum into two different regimes (i.e. the EIT and ATS). This finding should deepen our understanding of the behavior of a $\Lambda$-type three-level system.

The remainder of this paper is structured as follows: Sec.\ref{secII} includes two subsections. In Sec.\ref{secII.A}, we first present a theoretical model of the energy-level diagram considered. In Sec.\ref{secII.B}, we further consider the application of both quantum and classical methods to probe the spectrum of the sample. To quantify the accuracy of the measurement, we introduce the use of Fisher information for parameter estimation in the spectrum. The results presented in Sec.\ref{secII} are discussed, starting with an analysis of the Fisher information for the atomic system in Sec.\ref{secIII}. In Sec.\ref{secIII.A}, we analyze the behavior and the limitations of Fisher information in a two-level system, then further apply this to recognize the information contributed by the absorption. In Sec.\ref{secIII.B}, we consider the Fisher information in a $\Lambda$-type system. Carrying on the knowledge from the previous subsection, we further recognize that the composition of Fisher information in the $\Lambda$-type system can be seen as being contributed by the absorption and the transparency. We then analyze the demarcation between those two contributions and consider the limitations of the quantum advantage in a $\Lambda$-type system. At the end of this section, we discuss the impact of the quantum advantage on the decoherence rate and technical loss in a $\Lambda$-type system. Finally, a summary is presented in Sec.\ref{secIV}.

\section{\label{secII}Absorption spectroscopy scheme}

\subsection{\label{secII.A}Configuration of the atomic medium}


\begin{figure*}[!htbp]
\centering
\includegraphics[width=0.8\textwidth]{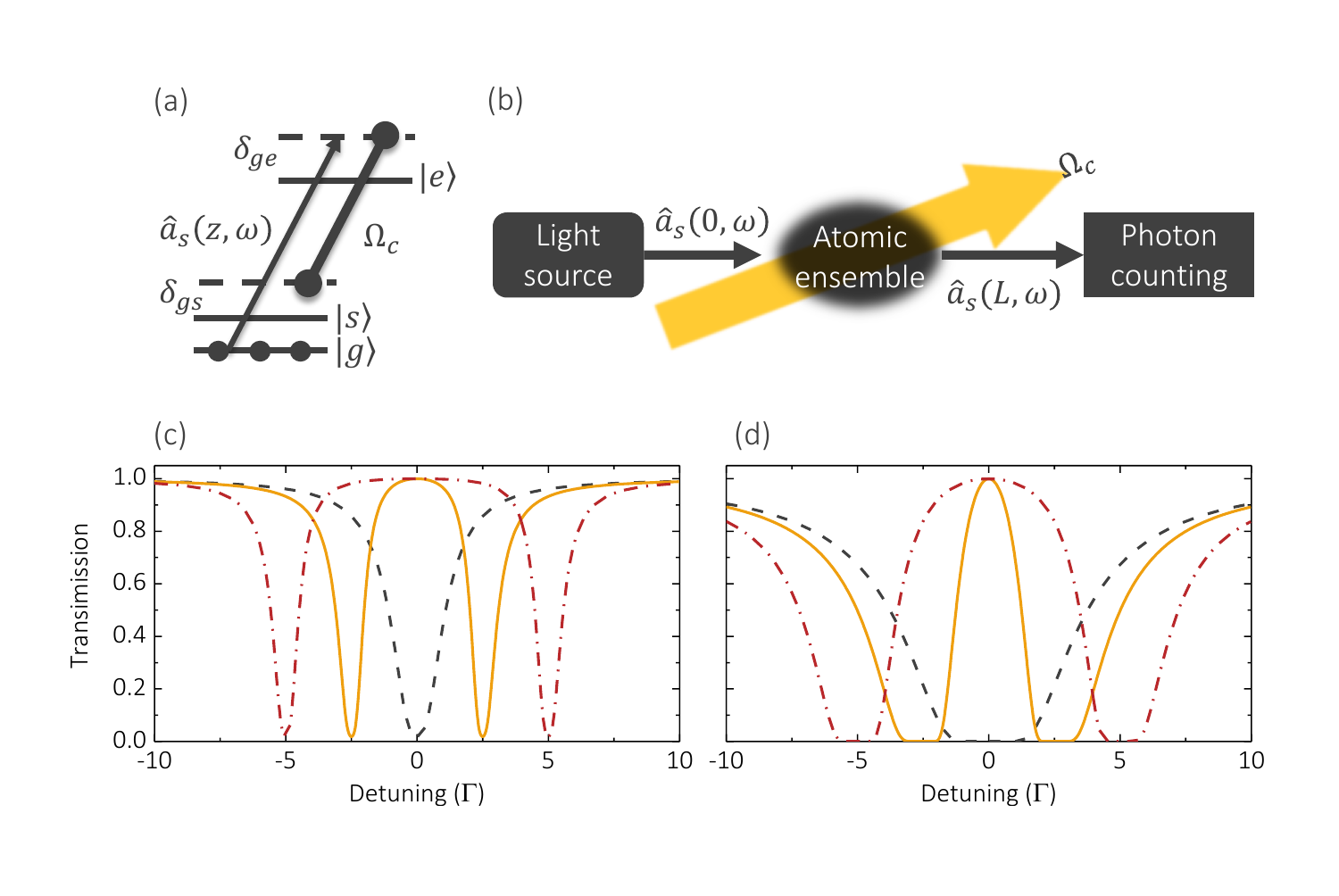}
\caption{(a) Schematic diagram of $\Lambda$-type atomic system; (b) scheme of absorption spectroscopy; (c) and (d) show the transmission spectra with optical depths of 4 and 40, respectively. Compare the behavior in different coupling strengths as indicated by the yellow line for 5 $\Gamma$, red dash-dotted line for 10 $\Gamma$, and black dashed line for 0 $\Gamma$.}
\label{Fig1}
\end{figure*}

We consider a $\Lambda$-type three-level atomic system with two ground states $|g\rangle$ and $|s\rangle$, and one excited state $|e\rangle$, as shown in Fig.\ref{Fig1} (a). A strong control field couples the transition between states $|s\rangle$ and $|e\rangle$. A weak quantized signal field $\hat{a}_s(z,\omega)$ drives the transition between states $|g\rangle$ and $|e\rangle$. We assume that the entire population is initially prepared in state $|g\rangle$. Under a weak signal perturbation, the Heisenberg-Langevin equations for atomic coherence operator $\hat{\sigma}_{ij}$ in the frequency domain can be written as
\begin{equation}
\begin{split}
-i\omega\hat{\sigma}_{ge}(z,\omega)=\frac{i}{2}\Omega_c\hat{\sigma}_{gs}(z,\omega)+\frac{i}{2}g_s\hat{a}_s(z,\omega)\\+\left(i\delta_{ge}-\frac{\gamma_{ge}}{2}\right)\hat{\sigma}_{ge}(z,\omega)+\hat{F}_{ge},
\end{split}
\label{si31eq}
\end{equation}
\begin{equation}
-i\omega\hat{\sigma}_{gs}(z,\omega)=\frac{i}{2}\Omega_c^*\hat{\sigma}_{gs}(z,\omega)+\left(i\delta_{gs}-\frac{\gamma_{gs}}{2}\right)\hat{\sigma}_{gs}+\hat{F}_{gs},
\label{si21eq}
\end{equation}
where $\delta_{ge}$ denotes the detuning of the signal field to $\omega_{ge}$. The two-photon detuning $\delta_{gs}$ is defined as $\delta_{gs}=\delta_{gs}-\delta_c$, where $\delta_c$ is the detuning of the control field; $\hat{F}_{gs}$ is the Langevin noise operator; and $g_s$ is the coupling constant for the interaction between the atoms and the signal field. $\gamma_{ge}=\Gamma$ is the spontaneous decay rate from $|e\rangle$ to the ground states and $\gamma_{gs}$ is the decoherence rate of the two ground states. The Maxwell-Schrodinger equation for the signal field reads
\begin{equation}
\left(-\frac{i\omega}{c}+\frac{\partial}{\partial z}\right)\hat{a}_s(z,\omega)=\frac{ig_sN}{c}\hat{\sigma}_{ge},
\label{MSE}
\end{equation}
\\
where $N$ is the atom number and $c$ is the speed of light in a vacuum. 

By solving the coupled equations, Eqs.\ref{si31eq}-\ref{MSE}, we get the field operator, after interacting with atomic medium of a length $L$,
\begin{equation}
\begin{aligned}
\hat{a}_s(L,\omega)=&\hat{a}_s(0,\omega)e^{-\Lambda(\omega)L}+\delta \hat{a}_s(L,\omega),\\
\delta \hat{a}_s(L,\omega)=&\int_0^L\frac{ig_sN}{c}\hat{f}(\omega)e^{\Lambda(\omega)(z'-L)}dz',
\label{a_eit}
\end{aligned}
\end{equation}
where
\begin{equation*}
\begin{aligned}
\Lambda(\omega)&=-\frac{i\omega}{c}+\frac{g_s^2Nd_{gs}}{2c\mathbf{D}(\omega)}\\
\mathbf{D}(\omega)&=d_{ge}d_{gs}+|\Omega_c/2|^2,\\
d_{ij}&=\gamma_{ij}/2-i(\omega+\delta_{ij}),\\
\hat{f}(\omega)&=(i\Omega_c\hat{F}_{gs}+d_{gs}\hat{F}_{ge})/\mathbf{D}(\omega), 
\end{aligned}
\end{equation*}
and parameter $\alpha=2NLg_s^2/\Gamma c$ is the optical depth of the medium. The first term of Eq.\ref{a_eit} represents the frequency response of the medium and the second term represents the noise due to the coupling of atoms to a radiative vacuum.

Since the phase information of cannot be analyzed in absorption spectroscopy, here we only consider the transmission of photons. To do so, we calculate the ratio between the input and output mean photon number. From Eq.\ref{a_eit}, we have
\begin{equation}
T(\delta_{ge})=\exp\left\{-2Re[\Lambda(0)]L\right\}.
\label{T_eit}
\end{equation}
Here, we have already considered the steady-state case of $\omega=0$ in Eq.\ref{T_eit}. In FIG.\ref{Fig1} (c) and (d), we based on Eq.\ref{T_eit}, show a plot of the spectra in the case of a relatively low ($\alpha=4$) and high ($\alpha=40$) optical depth with different coupling strengths. For the cases of $\Omega_c=0$ in both figures, since the signal field is considered as a weak field, the population of the atomic ensemble will not be excited to other states, as a result of which the system can be seen as a two-level system. For the cases of $\Omega_c\neq0$, we can observe that the so-call electromagnetically-induced-transparency (EIT) window appears.

The appearance of the transparency window is due to the Fano interference \cite{fano1961effects}, which describes the interference of the atomic polarization coherence. In the case of a relatively low coupling field, the Fano interference becomes stronger, inducing a destructive interference on the absorption that causes the transparency (EIT region). In the EIT region, the structure of the spectrum is dominated by transparency. When the $\Omega_c$ increases, the interference of the atomic coherence becomes weaker, resulting in the system being seen as having two absorption resonances, which are defined as the Autler-Townes Splitting (ATS) region. 

In general, the line of demarcation is considered to be around $\Omega_c\approx\Gamma$ for measuring the overlap of the two absorption resonances for atomic coherence \cite{abi2010electromagnetically,anisimov2011objectively}. However, in view of the transmission spectrum (or spectral line shape), it is hard to observe the demarcation between EIT and ATS when the optical depth parameter $\alpha$ must be considered. For instance, in the cases of $\Omega_c=5\Gamma$, as shown in FIG.\ref{Fig1} (c) and (d), the spectral line shape appears as two ATS absorption peaks for low $\alpha$, however, the line shape  appears to be  like EIT when $\alpha$ increased even though the condition of $\Omega_c=5\Gamma$  has already defined the region as ATS. Therefore, a grey area still exists when classifying the spectral line shape of the $\Lambda$-type system.

In the following section, we consider absorption spectroscopy, using both quantum and classical methods to probe the medium. Moreover, we introduce Fisher information to quantify the measurement. Since Fisher information possesses the information required for expressing the structure of the spectrum, it provides a way to classify the spectrum which is normally dominated by transparency or absorption from the perspective of the spectral line shape.

\subsection{\label{secII.B}Fisher information by photon counting}


Now consider the absorption spectroscopy scheme, as shown in FIG.\ref{Fig1} (b). The prepared light source sends the light field to the atomic medium for detection. For the measurement part of absorption spectroscopy, consider a means of simple detection by photon-number-counting. To estimate the lower bound of the variance of any parameter $\theta$ in the spectrum, we consider the Cram{\'e}r-Rao bound \cite{cramer1999mathematical},
\begin{equation}
\left \langle \left(\Delta\theta \right)^2 \right \rangle\geq\frac{1}{F(\theta)},
\label{CRB}
\end{equation}
where $F(\theta)$ is the Fisher information (FI) for estimating the information of parameter $\theta$ in a specific measurement scheme with a given quantum state of the probe light. Here we calculate FI by the form of 
\begin{equation}
F(\theta)=\sum_{n=0}^{\infty}\frac{1}{p(n|\theta)}\left(\frac{\partial}{\partial \theta}p(n|\theta)\right)^2,
\label{FI}
\end{equation}
where $p(n|\theta)$ is the conditional probability of measuring $n$ photons when the medium condition parameter is set at $\theta$. In our case, this parameter could be transmission $T$ or the frequency difference between the medium transition and the input photons $\delta_{ge}$.

To illustrate the quantum enhancement of parameter estimation by using nonclassical light, we calculate the FI in both cases, using the coherent state (classical case) and the single-photon-Fock-state (quantum case), to be the light source for estimating the parameters in the spectrum. In the following discussion, assume that the coupling detuning ($\Delta_{C}$) is set to 0 for focusing on the physical connotations. The technical loss $\eta$ and decoherence rate $\gamma_{gs}$ describe the experimental imperfection. Therefore, the relation of the input-output intensity is given by
\begin{equation}
I=I_0\eta T(\delta_{ge}),
\label{I_I0}
\end{equation}
where $I$ is the mean intensity of the output light and $I_0$ denotes the intensity of the input light. 

\subsubsection{Classical case}
In the case using a classical light source, the quantum state of light is considered as a coherent state $\left|\alpha\right\rangle$. According to Eq.\ref{a_eit}, Eq.\ref{T_eit} and Eq.\ref{I_I0}, the output photon number distribution as a function of transmission can be expressed by
\begin{equation}
p_c(n|T)=e^{-\eta T(\delta_{ge})|\alpha|^2}\frac{\eta^nT(\delta_{ge})^n|\alpha|^{2n}}{n!}.
\label{pcT}
\end{equation}
By using Eq.\ref{FI} and Eq.\ref{pcT}, it is easy to solve the FI for transmission estimation in the classical case as follows:
\begin{equation}
\mathcal{F}_c(T)=\frac{F_c(T)}{|\alpha|^2}=\frac{\eta}{T(\delta_{ge})}.
\label{FcT}
\end{equation}
In Eq.\ref{FcT}, we already consider the mean Fisher information $\mathcal{F}$ which is contributed by each photon. For frequency estimation, we apply variable transformation on Eq.\ref{FI} \cite{lehmann2006theory}, which results in 
\begin{equation}
F(\delta_{ge})=\left(\frac{\partial T}{\partial\delta_{ge}}\right)^2F(T).
\label{FI1}
\end{equation}
Notice that, Eq.\ref{FI1} shows that the FI for frequency estimation is obtained by the multiplication of FI for transmission estimation with the factor of variable transformation $(\partial T/\partial\delta_{ge})^2$, which is the square of the differential of the spectrum which means that the FI for frequency estimation has the ability to react to the structure of the spectrum. By considering Eq.\ref{FI1}, the FI for frequency estimation in the classical case without decoherence rate is given by
\begin{equation}
\begin{aligned}
\mathcal{F}_c(\delta_{ge})=&\eta\frac{64\alpha^2\Gamma^4\delta_{ge}^2(|\Omega_c|^4-16\delta_{ge}^4)^2}{\left [4\Gamma^2\delta_{ge}^2+(|\Omega_c|^2-4\delta_{ge}^2)^2  \right ]^4}\\
\times& \exp\left[-\frac{4\alpha\Gamma^2\delta^2_{ge}}{4\Gamma^2\delta_{ge}^2+(|\Omega_c|^2-4\delta_{ge}^2)^2}\right].
\end{aligned}
\label{Fcd}
\end{equation}

\subsubsection{Quantum case}
For the quantum case, we use a single-photon source to probe the atomic medium. In this case, the output only contains a single-photon (photon passes through the atomic medium) or a vacuum state (photon is absorbed by the atomic medium). Thus, the output photon number distribution can be given by
\begin{equation}
p_q(n|T)=\left\{\begin{matrix} 1-\eta T, n=0
\\ 
\eta T, n=1
\end{matrix}\right..
\label{pqT}
\end{equation}
Based on Eq.\ref{pqT}, the FI for transmission estimation in the quantum case can be solved as
\begin{equation}
\mathcal{F}_q(T)=\frac{\eta}{T(1-\eta T)}.
\label{FqT}
\end{equation}
Further, the frequency estimation of FI also can be obtained by considering Eq.\ref{FI1} given as in the following equation:
\begin{equation}
\begin{aligned}
\mathcal{F}_q(\delta_{ge})=&\eta\frac{64\alpha^2\Gamma^4\delta_{ge}^2(|\Omega_c|^4-16\delta_{ge}^4)^2}{\left [4\Gamma^2\delta_{ge}^2+(|\Omega_c|^2-4\delta_{ge}^2)^2  \right ]^4}\\
\times&\left\{\exp\left[\frac{4\alpha\Gamma^2\delta^2_{ge}}{4\Gamma^2\delta_{ge}^2+(|\Omega_c|^2-4\delta_{ge}^2)^2}\right]-\eta\right\}^{-1},
\end{aligned}
\label{Fqd}
\end{equation}
where $\gamma_{gs}$ is set at 0 to simplify the equation. 
\subsubsection{Quantum-enhanced factor}

To show the enhancement between the quantum case and the classical method, here we define the quantum-enhanced factor (QEF) of $Q=\mathcal{F}_q/\mathcal{F}_c$ \cite{england2019quantum,allen2020approaching,whittaker2017absorption}. According to Eq.\ref{FI1}, the QEF always has the same behavior even when the estimation parameter is changed. Considering Eq.\ref{FcT} and Eq.\ref{FqT}, the QEF in our case can be expressed as 

\begin{equation}
Q=\frac{\mathcal{F}_q}{\mathcal{F}_c}=\frac{1}{1-\eta T}.
\label{QEF}
\end{equation}

\begin{figure}[t]
\centering
\includegraphics[width=0.48\textwidth]{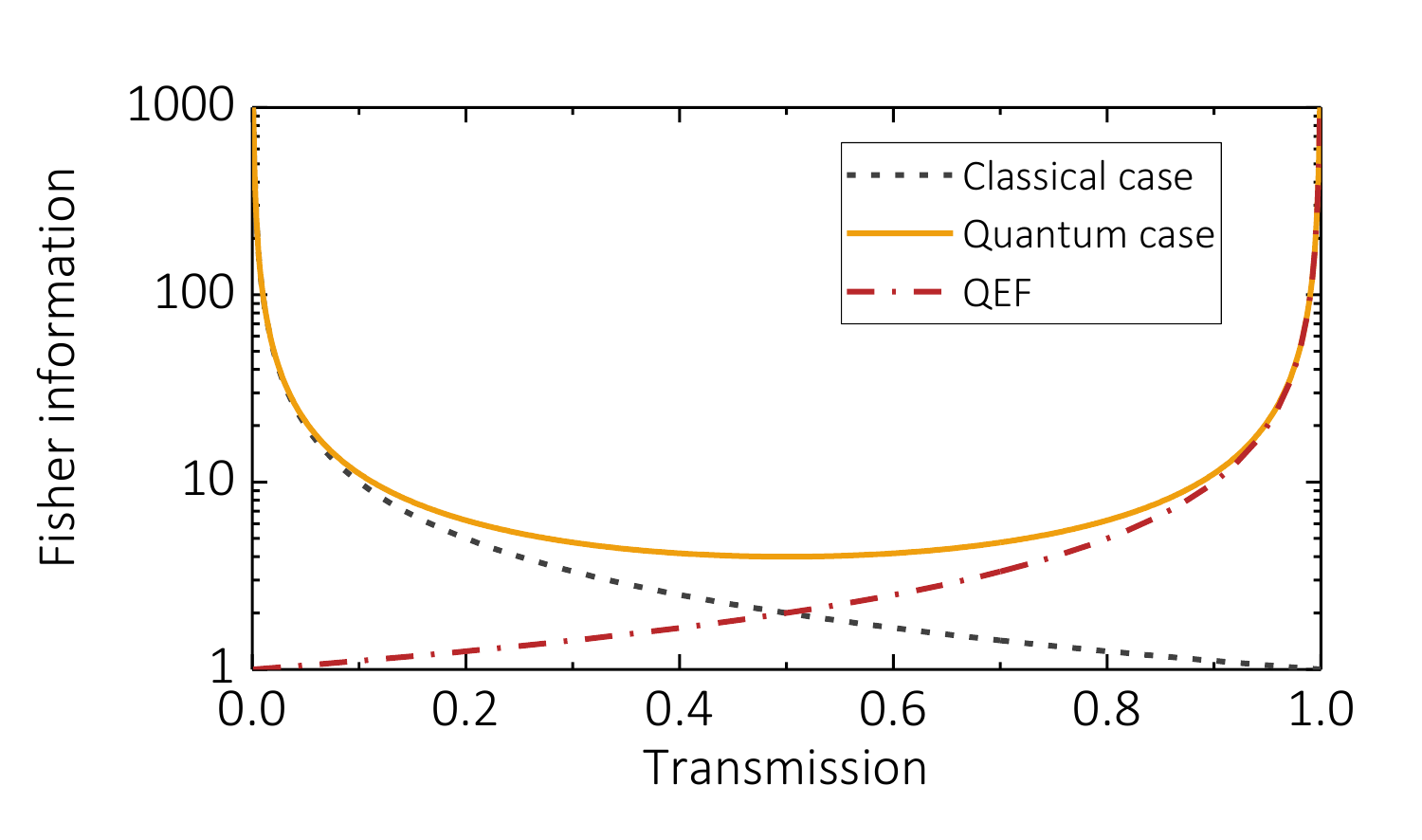}
\caption{Comparison of FI for transmission estimation in both quantum and classical cases. The yellow solid line denotes the quantum case, the black dashed line is the classical case, and the red dash-dotted line represents the quantum-enhanced factor.}
\label{Fig4}
\end{figure}

\begin{figure*}[!htbp]
\centering
\includegraphics[width=0.9\textwidth]{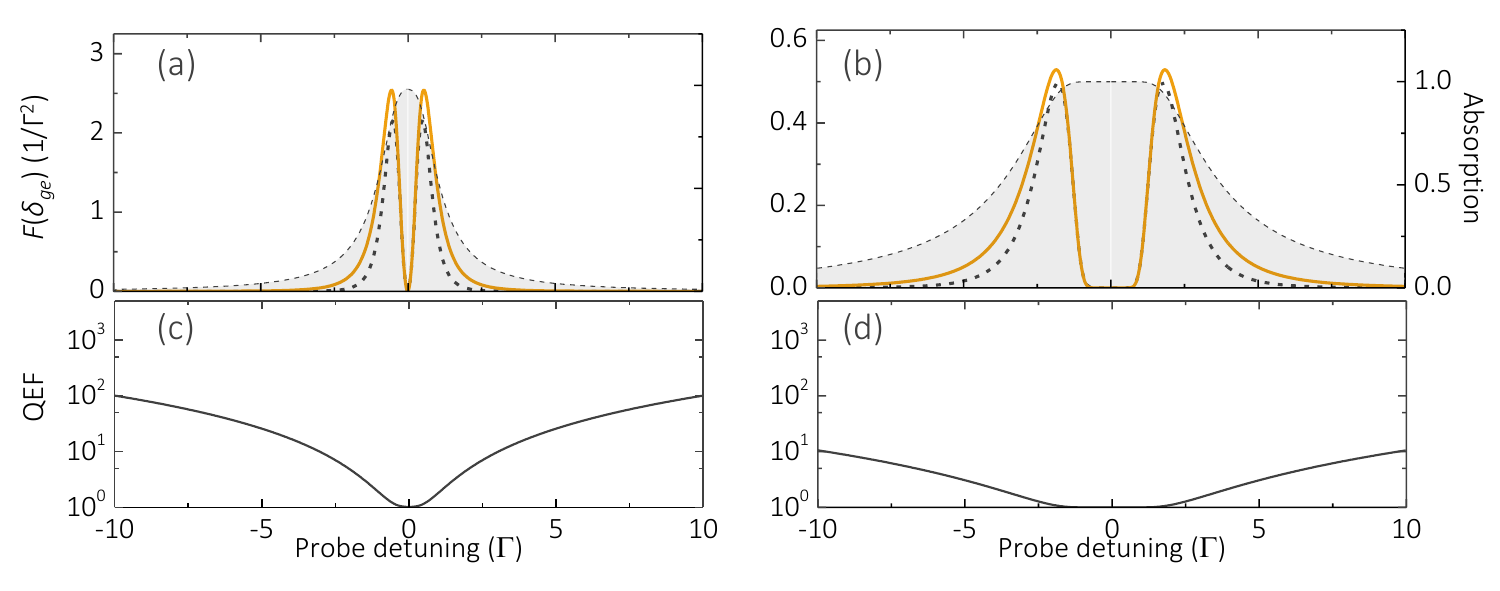}
\caption{The FI in the quantum (yellow solid line) and classical (black dashed line) cases in a two-level system without technical loss ($\eta=1$). All results are calculated by using Eq.\ref{Fcd}, Eq.\ref{Fqd}, and Eq.\ref{QEF}. The gray area represents the absorption of the atomic medium. The optical depth is set to $\alpha=4$ in the cases of (a) and (c), and to $\alpha=40$ for the cases of (b), and (d). In (a) and (b), we show the FI for frequency estimation as a function of detuning. The quantum enhancement is shown in (c) and (d).}
\label{Fig5}
\end{figure*}

FIG.\ref{Fig4} shows the FI for transmission estimation in both the quantum and classical cases and the QEF without a technical loss ($\eta=1$). Here we can observe that, in the low transmission region ($T\approx0$), $\mathcal{F}_q\approx\mathcal{F}_c$ and both approaches to $\infty$. This behavior can be understood by the following physical picture: since the photon number distribution of the output in both the quantum and classical cases is dominated by the vacuum state when $T\approx0$, the contrast between the input and output photon number distribution is close. Therefore, the value of FI is almost the same in both cases, which results in $Q\approx1$ in the low transmission region. For the high transmission region ($T\approx1$). Since the photon number distribution of the classical (coherent) state has a larger variance than in the quantum case (single-photon Fock state) of defined photon numbers, the measurement obtained by photon counting in a classical case  has higher fluctuation in the high transmission region, and therefore we can observe a high quantum enhancement in the high transmission region. An interesting situation occurs in the quantum case at $T=0.5$. In this case, the output photon number distribution becomes well-distributed which corresponds to the largest variance of the output photon number distribution of the quantum case, therefore, the FI has the lowest value at this point. 

FIG.\ref{Fig4} shows an important result that is only visible in the high transmission region (i.e. low absorption). The quantum enhancement is obvious. This property will strongly encourage the behavior of the FI for spectrum estimation, furthermore, this also implies that the single-photon state does not actually produce a clear advantage for detecting the spectrum of a simple absorption structure. However, from another angle, the single-photon state may be more suitable for detecting a spectrum with a transparent window structure, e.g. the EIT spectrum. Additionally, the QEF discussed here only considers the case where the parameters are the same for both the quantum and classical method. However, it is also possible to find the optimal conditions for both schemes to maximize the FIs, for a further more equitable comparison of the QEF \cite{allen2020approaching}. In particular, for frequency estimation in an atomic system, the numerous parameter space opens up the rich phenomena of FI, as discussed in the next section.


\section{Fisher information spectrum}\label{secIII}

In this section, we discuss the FI for estimating the frequency in both two-level and $\Lambda$-type spectra. The structure and the limitations of FI for frequency estimation are analyzed. The  structural analysis focuses on  finding  optimal detuning for the maximum FI at a fixed optical depth. By measuring the trace of those maximum FIs, we can understand the structure of the FI spectrum. In terms of the limitations of QEF, the analysis is based on finding the optimal optical depth for a maximum FI at fixed detuning. Using the analytical method described above, we can further obtain the QEF for both two-level and $\Lambda$-type systems.


\subsection{Two-level system}\label{secIII.A}
Starting with the conditions of $\Omega_c=0$ and $\eta=1$ which  simplify the system into two-level system. Since the two-level system does not couple to the $\left|s\right\rangle$ state, the FI is independent on $\gamma_{gs}$. In this case, the variable parameter of the spectral line shape is the optical depth $\alpha$ only. Here we compare the behavior of FI for frequency estimation, in both the quantum and classical cases, in a two-level system with different optical depths, as shown in FIG.\ref{Fig5}. 

In the frequency estimation, we consider Eq.\ref{FI1}, Eq.\ref{Fcd} and Eq.\ref{Fqd} for calculating $\mathcal{F}_c(\delta_{ge})$ and $\mathcal{F}_q(\delta_{ge})$. Notice that, due to the structure of the two-level absorption spectrum, we can see that the FI for both cases are always zero at the point of resonance [due to  $(\partial T/\partial\delta_{ge})^2|_{\delta_{ge}=0}=0$]. This naturally results is no information on the resonance center, for  both the quantum and classical cases in a two-level system, which means that we  can never precisely estimate the frequency of the resonance in the two-level system by photon-counting measurement. On the other hand, unlike the case of transmission estimation \cite{allen2020approaching}, since the high $\alpha$ flattens the curve of the absorption spectrum around resonance, the FI is eliminated in the high $\alpha$ case within a broader range. 

\begin{figure*}[!htbp]
\centering
\includegraphics[width=0.9\textwidth]{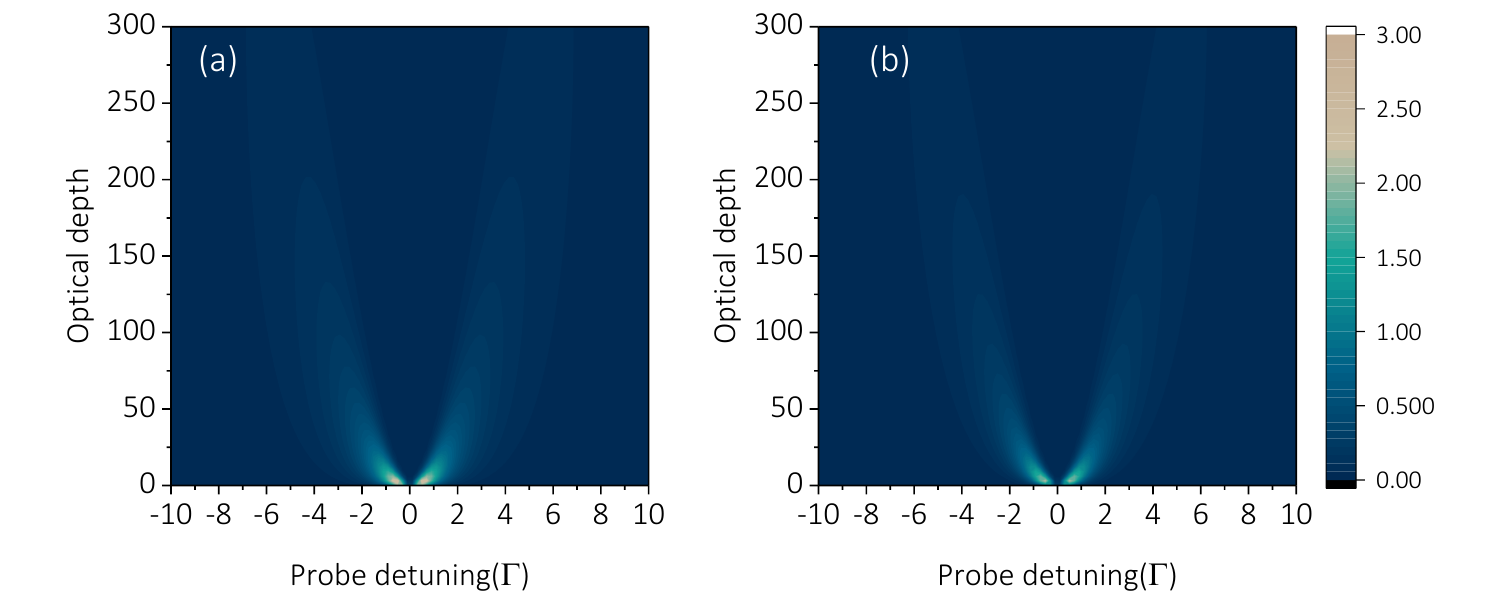}
\caption{Density plots of Fisher information for the (a) quantum and (b) classical cases in a two-level system. The unit of the color bar (FI) is $1/\Gamma^2$.}
\label{Fig6}
\end{figure*}

Although the information cannot be analyzed within the high absorption region, we can still find two main peaks which dominate the structure of FI. Now we discuss the behavior of the FI peaks in the spectra. In FIG.\ref{Fig5} (a) and (b), we can observe that the distance between the two peaks is dependent on the optical depth. Since the high $\alpha$ broadens in the high absorption range, the spacing between two peaks increases as the optical depth increases, and vice versa. FIG.\ref{Fig6} shows a density plot of FI from which we can observe the continuous behavior of FI. 

\begin{figure}[t]
\centering
\includegraphics[width=0.48\textwidth]{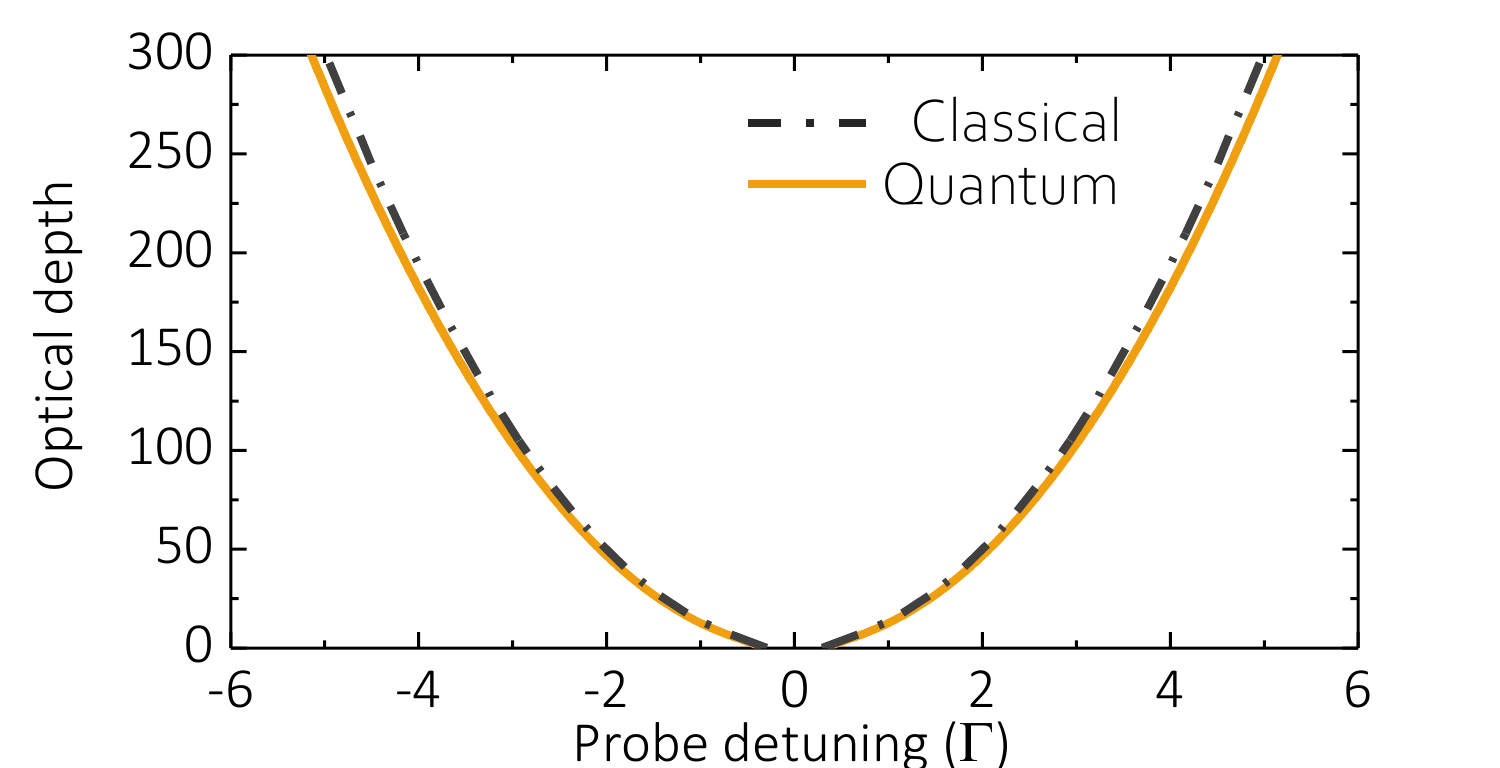}
\caption{Tracing the maximum Fisher information in a phase diagram for the quantum (yellow solid line) and classical (black dashed line) cases. The curves represent the relationship between the optical depth and the probe detuning based on the maximum Fisher information.}
\label{Fig7}
\end{figure}

\begin{figure}[t]
\centering
\includegraphics[width=0.48\textwidth]{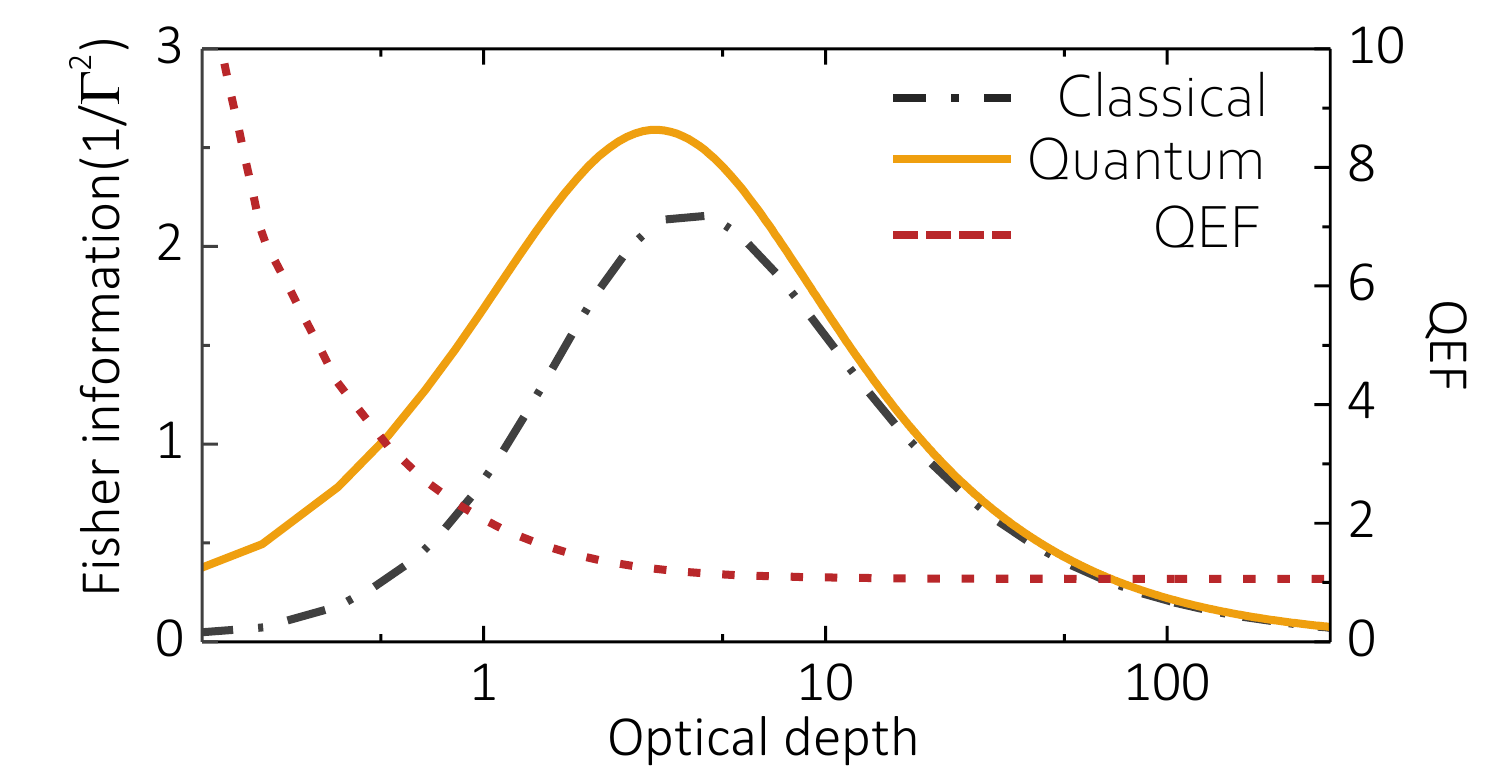}
\caption{Comparison of the Fisher information (left axis) in the quantum (yellow solid line) and classical (black dash-dotted line) cases when the detuning is set to the optimal conditions of $\delta^T_{q,opt}$ and $\delta^T_{c,opt}$. On the right axis, we show the QEF (red dashed line) between the FI.}
\label{Fig8}
\end{figure}

As can be seen in FIG.\ref{Fig6},  the typical behavior is quite similar in both the quantum and classical cases. However, we should note the difference in the position of the optimal FI at a fixed $\alpha$. The reason for this is that, since the quantum case has a quantum advantage in the low absorption region, the optimal value of $\mathcal{F}_q(\delta_{ge})$ will appear in the area of farther detuning from the resonance center than in the classical case. This can also be observed in FIG.\ref{Fig5} (a) and (b). 

In some experimental cases, the sample may not allow the optical depth to change (i.e. with a fixed $\alpha$), therefore, we can choose an optimal detuning then find the corresponding optimal FI. These optimal FI can be seen as a representation of the structure of the FI spectrum. Furthermore, this analysis can also help us to compare the QEF in optimal conditions to further show the limitation of the quantum advantage. In order to find those optimal detuning of $\delta^T_{q,opt}$ and $\delta^T_{c,opt}$ for both quantum and classical cases, respectively, we trace the maximum value of FI in FIG.\ref{Fig6} at a fixed $\alpha$; the results are shown in FIG.\ref{Fig7}. Using FIG.\ref{Fig7}, compare $\mathcal{F}_c(\delta^T_{c,opt})$ and $\mathcal{F}_q(\delta^T_{q,opt})$ in different $\alpha$, as shown in FIG.\ref{Fig8}. An important property is that even though the detuning is chosen to match the optimal FI, it still has no quantum advantage in the high $\alpha$ region ($\alpha\gg1$). To increase the QEF, we should prepare a low $\alpha$ sample. However, for the absolute value of FI, we can see there exists an optimal optical depth to maximize the FI in both cases. 

When the experiential conditions allows us to vary the optical depth of the sample, it is possible to find the maximum FI in a two-level system and to further estimate a specific frequency for the spectrum. To do so, first solved the equations of $\partial\mathcal{F}_c(\delta_{ge})/\partial \alpha=0$ and $\partial\mathcal{F}_q(\delta_{ge})/\partial \alpha=0$ in order to find the optimal optical depth for the classical and quantum cases of $\alpha^T_{c,opt}$ and $\alpha^T_{q,opt}$, respectively. The result is given by   

\begin{equation}
\alpha^T_{c,opt}=2\left(\frac{4\delta_{ge}^2}{\Gamma^2}+1\right),
\label{alpha_copt}
\end{equation}
and 
\begin{equation}
\alpha^T_{q,opt}=\alpha^T_{c,opt}\left[1+\frac{1}{2}W(-\frac{2\eta}{e^2})\right],
\label{alpha_qopt}
\end{equation}
where $W(x)$ is the Lambert $W$-function. By inserting Eq.\ref{alpha_copt} and Eq.\ref{alpha_qopt} into Eq.\ref{Fcd} and Eq.\ref{Fqd}, respectively, we can find the optimal FI for both the quantum and classical cases at an optical depth of $\alpha^T_{c,opt}$ and $\alpha^T_{q,opt}$, as follows:
\begin{equation}
\mathcal{F}_{q,\alpha}^{T}=-\frac{128\delta_{ge}^2W(-2\eta/e^2)[1+\frac{1}{2}W(-2\eta/e^2)]}{(4\delta_{ge}^2+\Gamma^2)^2},
\label{Fqopt_a}
\end{equation}
and 
\begin{equation}
\mathcal{F}_{c,\alpha}^{T}=\frac{256\eta\delta_{ge}^2}{e^2(4\delta_{ge}^2+\Gamma^2)^2},
\label{Fcopt_a}
\end{equation}
where $\mathcal{F}_{c,\alpha}^{T}$ and $\mathcal{F}_{q,\alpha}^{T}$ denotes the FI in a two-level system at the optimal optical depth for the classical and quantum cases, respectively. 

An important feature is shown in Eq.\ref{Fqopt_a} and Eq.\ref{Fcopt_a}. Here, we can observe when the optical depth is chosen at optimal conditions, the maximum values of $\mathcal{F}_{c,\alpha}^{T}$ and $\mathcal{F}_{q,\alpha}^{T}$ will always happen at the same detuning of $\pm\Gamma/2$, which is half of the absorption linewidth, and furthermore, be independent of the technical loss $\eta$. 

By using the results above, we can further find the corresponding optical depths in both cases. By substituting $\delta_{ge}=\pm\Gamma/2$ into Eq.\ref{alpha_copt} and Eq.\ref{alpha_qopt}, we find that the best  optical depth conditions for the classical and quantum cases are given by 4 and $4\left[1+1/2W(-2\eta/e^2)\right]$, respectively. In addition, this also leads to obtaining the maximum values of FI in a two-level system in both cases, which are $\mathcal{F}^T_{c,max}=16\eta/e^2\Gamma^2$ and $\mathcal{F}^T_{q,max}=-8W(-2\eta/e^2)(1+W(-2\eta/e^2)/2)/\Gamma^2$. Moreover, we can find the QEF in the case of
\begin{equation}
Q^{T}_{\alpha,opt}=-\frac{e^2}{2\eta}W(-\frac{2\eta}{e^2})\left[1+\frac{1}{2}W(-\frac{2\eta}{e^2})\right].
\label{Q_opt}
\end{equation}
A few important results from the above discussion are presented as follows: there is a fundamental limit to frequency estimation in the absorption spectrum when using photon-counting measurement. For the perfect case, of $\eta=1$, the FI of the quantum and classical cases are limited by  $\mathcal{F}^T_{c,max}\approx2.16/\Gamma^2$ and $\mathcal{F}^T_{q,max}\approx2.6/\Gamma^2$, respectively. The QEF is also limited to  $Q^{T}_{\alpha,opt}=e^2W(-2/e^2)(1+W(-2/e^2))/2\approx1.2$, which is consistent with the results obtained in Ref.\cite{allen2020approaching}. Corresponding to the practical experiment, the CRB provides a more intuitive relation for estimating the variance of observable. Consider Eq.\ref{CRB}, the lower bounds of the variance for the frequency estimation in a two-level system are given by $\langle(\Delta\delta_{q,ge})^2\rangle\geq0.621\Gamma^2$ and $\langle(\Delta\delta_{c,ge})^2\rangle\geq0.679\Gamma^2$, respectively. 

The FI analysis of a two-level system provides a model to identify the contribution of the absorption to the behavior of FI. We find the limitation and structure of FI in the two-level system in both quantum and classical methods. Interestingly, when the system becomes a $\Lambda$-type system ($\Omega_c\neq0$), the spectrum is not only derived from the absorption but also the transparency, which makes it become a mixture of ATS and EIT. 

In the next section, we analyze a $\Lambda$-type system to recognize and classify the composition of FI then further find the limitation of QEF in a $\Lambda$-type system.

\begin{figure*}[!htbp]
\centering
\includegraphics[width=0.9\textwidth]{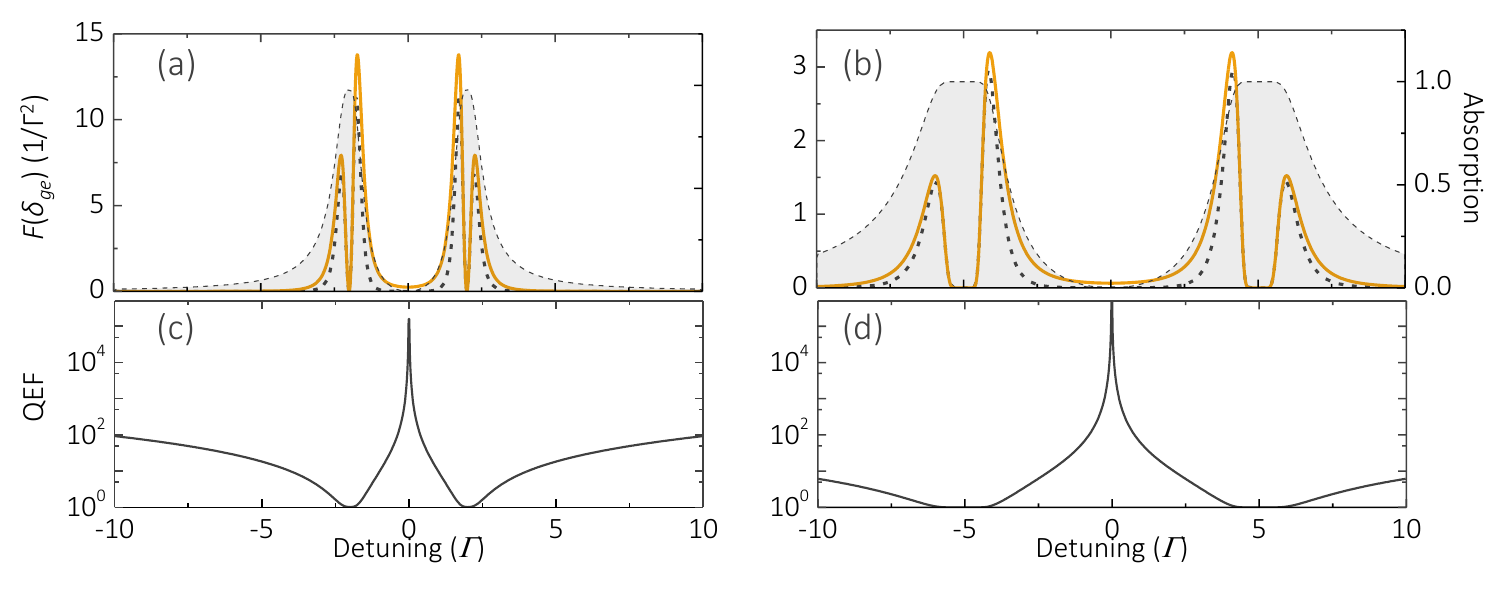}
\caption{Demonstration of the absorption-dominated Fisher information spectrum. The FI in quantum (yellow) and classical (black) cases in a $\Lambda$-level system without technical loss ($\eta=1$). All results are calculated by using Eq.\ref{Fcd}, Eq.\ref{Fqd}, and Eq.\ref{QEF}. The gray area represents the absorption of the atomic medium. The optical depth is set $\alpha=4$ and $\Omega_c=4\Gamma$ in the cases of (a) and (c), $\alpha=40$ and $\Omega_c=10\Gamma$ for the cases of (b), and (d). In (a) and (b), we show the FI for frequency estimation as a function of detuning. The quantum-enhanced factors for (a) and (b) are shown in (c) and (d), respectively.}
\label{Fig9}
\end{figure*}

\begin{figure*}[!htbp]
\centering
\includegraphics[width=0.9\textwidth]{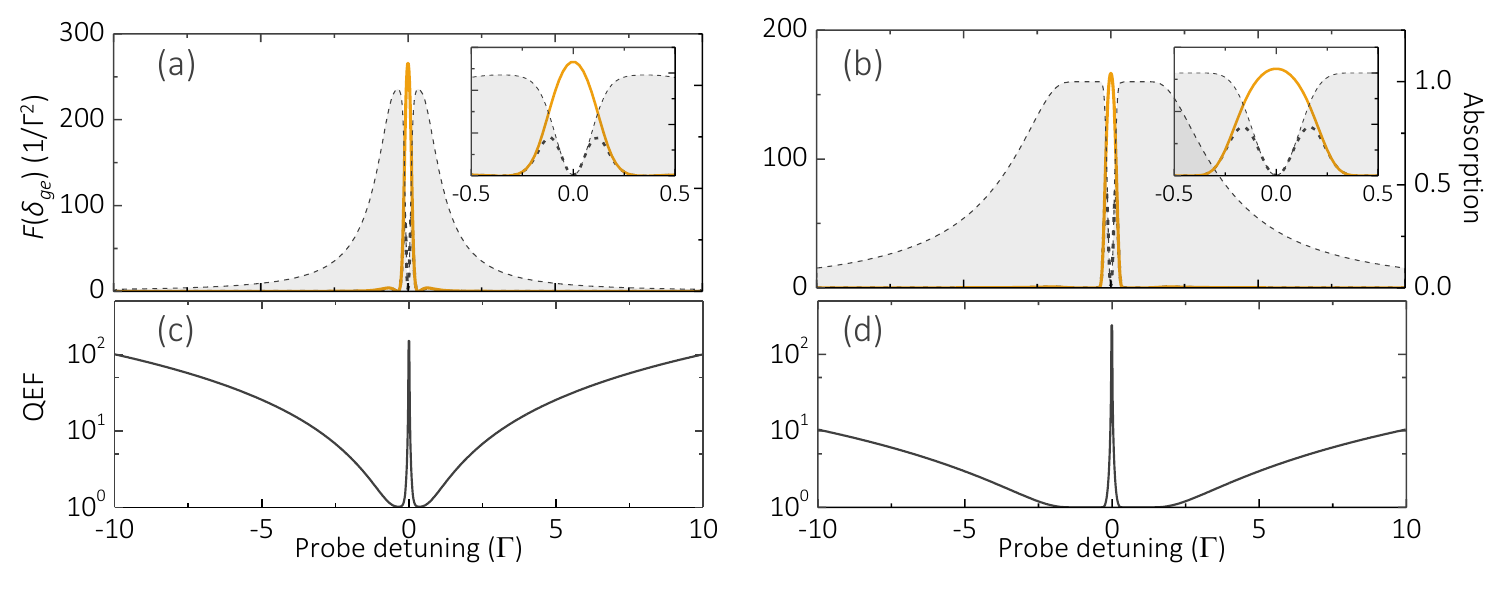}
\caption{Demonstration of the transparency-dominated Fisher information spectrum. The FI in quantum (yellow solid line) and classical (black dashed line) cases in a $\Lambda$-level system without technical loss ($\eta=1$). The gray area represents the absorption of the atomic medium. The optical depth is set to  $\alpha=4$ and $\Omega_c=0.7\Gamma$ in the cases of (a) and (c), $\alpha=40$ and $\Omega_c=1.4\Gamma$ for the cases of (b) and (d). In (a) and (b), we show the FI for frequency estimation as a function of detuning. The quantum-enhanced factors for (a) and (b) are shown in (c) and (d), respectively.}
\label{Fig10}
\end{figure*}

\subsection{$\Lambda$-type system}\label{secIII.B}

For a $\Lambda$-type system, the coupling fields become nonzero ($\Omega_c\neq0$). As for the two-level system, we first consider Eq.\ref{FcT} and Eq.\ref{FqT} in the perfect case when $\eta=1$ and $\gamma_{gs}=0$, to illustrate the physical concept; see section III.B. 1-4. In sections III.B. 5 and 6, we discuss the impact of a finite decoherence rate and the technical loss.



\subsubsection{Strong $\Omega_c$}
We start with a relatively strong $\Omega_c$, so that the system is close to the ATS region. The FI for both the quantum and classical cases is shown in FIG.\ref{Fig9}. In FIG.\ref{Fig9},  we can observe that the spectrum is dominated by two ATS peaks in FIG.\ref{Fig9} (a) and (b). In terms of the behavior of FI, there are two main FI peaks around each absorption peak, corresponding to the behavior of the two-level system (see FIG.\ref{Fig5}), in both the quantum and classical cases. However, since the spectral line shapes of the two ATS peaks are not symmetrical, the two main peaks of the FIs are not equal in height. Nevertheless, this case shows that when the system is close to the ATS region, the behavior of FI is dominated by the absorption of the spectral line shape of the system. Here we defined the FI spectrum as \emph{absorption-dominated}.

\subsubsection{Weak $\Omega_c$}
As discussed above, in the two-level case there is no information at the resonance center, in either the classical or the quantum cases. In the $\Lambda$-type system, however, there is a big difference between the classical and quantum cases. For the classical case,  the result is similar to that obtained with the two-level system. Consider Eq.\ref{FI1}, the factor of $(\partial T/\partial \delta_{ge})^2$ is equal to 0 at the resonance center. Combining this with $\mathcal{F}_c(T(0))=1$, we get the Fisher information for frequency estimation which is equal to 0 at the resonance center. In contrast, the FI in the quantum case provides infinite information for $\mathcal{F}_q(T(0))=\infty$, which results in an undefined value at the resonance center. To estimate the FI in the condition of $\delta_{ge}\rightarrow0$, we take the limit of
\begin{equation}
\mathcal{F}^{\Lambda}_{q,res}\equiv\lim_{\delta_{ge}\rightarrow0}\mathcal{F}_q(\delta_{ge})=\frac{16\alpha\Gamma^2}{|\Omega_c|^4}=\frac{16ln2}{\Delta\omega_{EIT}^2}. 
\label{FI_lim}
\end{equation}
where $\mathcal{F}^{\Lambda}_{q,res}$ represents the Fisher information that is contributed to by the EIT-transparency-window and $\Delta\omega_{EIT} $ is the bandwidth of the window of $\sqrt{ln2/\alpha}|\Omega_c|^2/\Gamma$ \cite{tsai2020quantum}. Interestingly, even though the FI for transmission estimation in the quantum case provides infinite information on the resonance center, the FI for frequency estimation is still limited by the structure of the EIT spectrum and gives a finite value. On the other hand, Eq.\ref{FI_lim} shows that the FI in the quantum case for estimating the frequency of the resonance center is determined by the EIT bandwidth directly. A narrower bandwidth provides higher information and vice versa.

Since the parameters of $\Omega_c$ and $\alpha$ are not limited, theoretically, this implies that $\mathcal{F}^{\Lambda}_{q,res}$ has the ability to approach infinity in the quantum case. In contrast, the FI contributed by the absorption has a limitation as $\alpha$ increases (see FIG.\ref{Fig8},). Therefore, we expect there to be a condition that makes $\mathcal{F}^{\Lambda}_{q,res}$ dominate the composition of FI in the quantum case. 

In FIG.\ref{Fig10}, it can be seen that the optical depths are the same in (a) and (b) as in FIG.\ref{Fig9} but they operate at a relatively weak $\Omega_c$, which means that the system is close to the EIT region. This condition leads it an increase in $\mathcal{F}^{\Lambda}_{q,res}$ and further exceeds the information that is contributed by absorption. Moreover, we can see that $\mathcal{F}^{\Lambda}_{q,res}$ dominates the FI spectrum. Therefore, in this case, we define this FI spectrum as \emph{transparency-dominated}.


\subsubsection{Transition between absorption- and transparency-dominated regions in the $\Lambda$-type spectrum}

\begin{figure*}[t]
\centering
\includegraphics[width=0.9\textwidth]{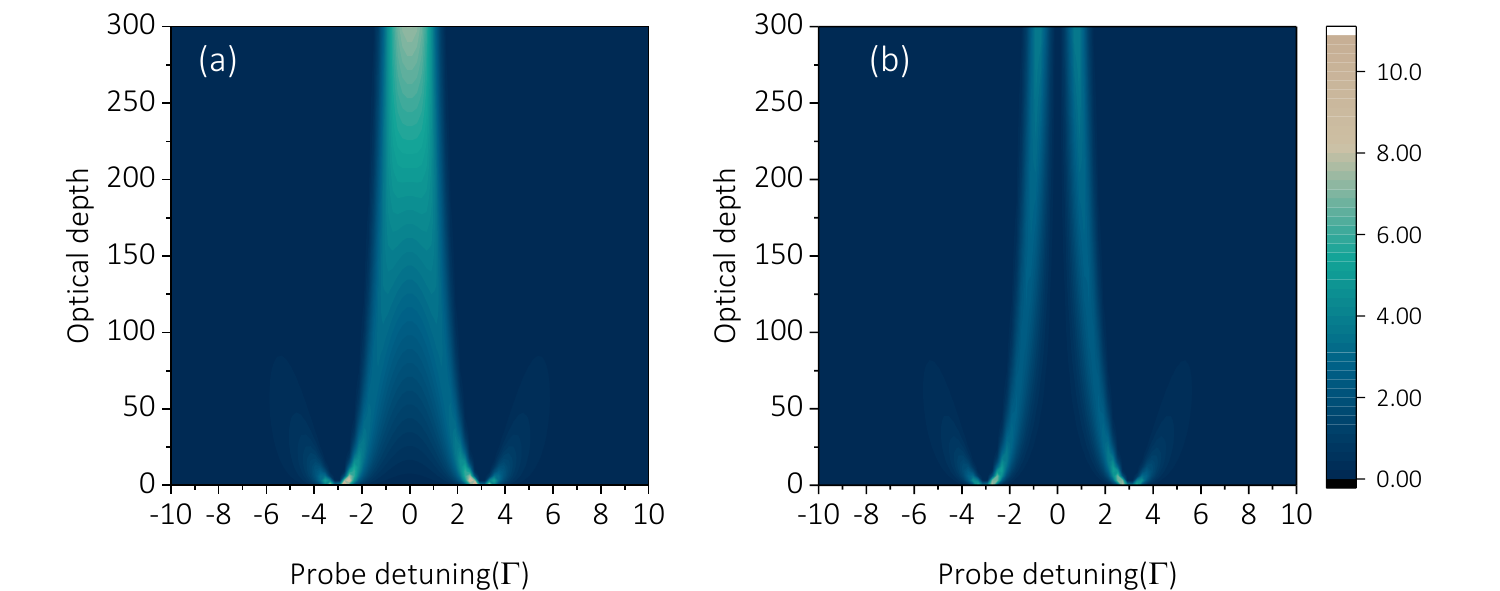}
\caption{The density plot of Fisher information for the (a) quantum and (b) classical cases in the $\Lambda$-type system at a fixed $\Omega_c=6 \Gamma$. The unit of the color bar is $1/\Gamma^2$.}
\label{Fig11}
\end{figure*}

An examination of FIG.\ref{Fig9} and FIG.\ref{Fig10} shows the existence of two regions where absorption and transparency dominate the FI spectrum in a $\Lambda$-type system. We are interested in the transition between those regions. In Eq.\ref{FI_lim}, with a fixed $\Omega_c$, the $\mathcal{F}^{\Lambda}_{q,res}$ increases as $\alpha$ increases. Thus, by comparing the FI from the absorption and $\mathcal{F}^{\Lambda}_{q,res}$, it is possible to find a demarcation point for distinguishing the class of the spectrum. FIG.\ref{Fig11}, shows a density plot of FI in a $\Lambda$-type system at a fixed $\Omega_c$ which illustrates the behavior of FI on a continuum.

\begin{figure}[t]
\centering
\includegraphics[width=0.48\textwidth]{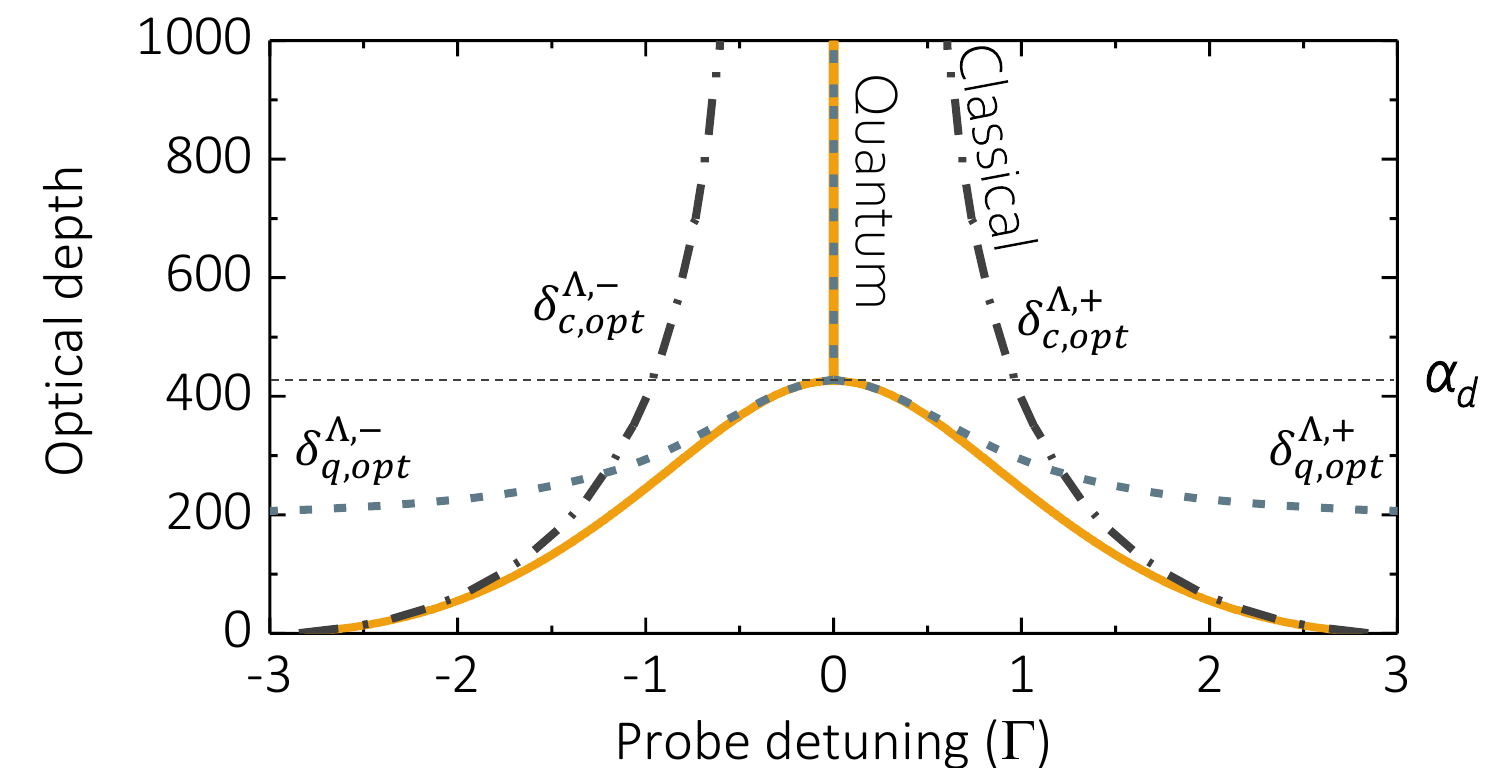}
\caption{Tracing the maximum Fisher information in the quantum (yellow solid line) and classical (black dash-dotted line) case. The blue dotted line denotes the approximate solution of Eq.\ref{d_ge_opt}.}
\label{Fig12}
\end{figure}

\begin{figure}[t]
\centering
\includegraphics[width=0.48\textwidth]{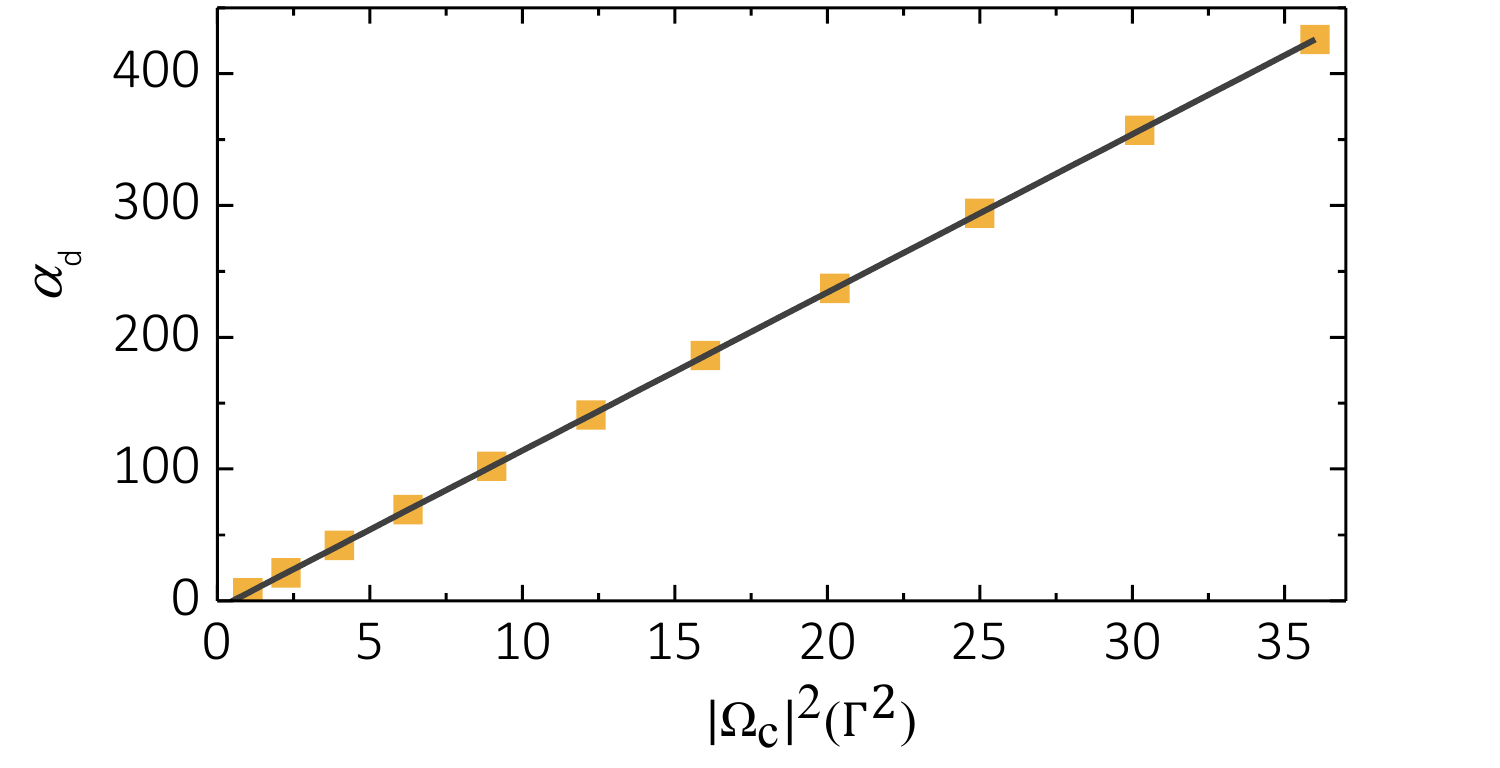}
\caption{The relation between the demarcation point $\alpha_d$ and the coupling strength $|\Omega_c|^2$. The yellow square denotes the results of the numerical calculation and the black line shows the results based on Eq.\ref{alpha_d}.}
\label{Fig13}
\end{figure}

Before further analysis of the class transition in the spectrum, we can clearly see that there is once again a big difference in behavior between the quantum and classical cases. In the quantum case [FIG.\ref{Fig11} (a)], following the increase in $\alpha$, the FI from the case where the two-main-peaks-dominated to the case of $\mathcal{F}^{\Lambda}_{q,res}$-dominated gradually. However, that does not happen in the classical case [FIG.\ref{Fig11} (b)] since the high transmission does not gain the information.

Let us return to the problem of the transition of the spectrum class. In order to find the demarcation, we trace the maximum value of FI in FIG.\ref{Fig11} at a fixed $\alpha$, as shown in FIG.\ref{Fig12}. The quantum case demonstrates an interesting phenomenon, that there is a clear demarcation point at a specific optical depth of $\alpha_{d}$, which represents the transition from an absorption-dominated to transparency-dominated spectral-line-shape. In order to analytically solve for the demarcation point, $\alpha_d$, we first expand Eq.\ref{Fqd} at $\delta_{ge}=0$, since the demarcation point occurs at  $\delta_{ge}=0$. Now we have

\begin{equation}
\begin{aligned}
\mathcal{F}_{q}(\delta_{ge})&=\frac{16\alpha\Gamma^2}{|\Omega_c|^4}-\frac{32\alpha\Gamma^2}{|\Omega_c|^8}[(6+\alpha)\Gamma^2-12|\Omega_c|^2]\delta_{ge}^2\\
&+\frac{64\alpha\Gamma^2}{3\Omega_c^{12}}
\begin{Bmatrix}
[72+\alpha(\alpha+24)]\Gamma^4\\ 
-48(\alpha+6)\Gamma^2\Omega_c^2+228\Omega_c^4
\end{Bmatrix}\delta_{ge}^4\\
&+O(\delta_{ge}^8). 
\label{Fqd_taylor}
\end{aligned}
\end{equation}

Notice that the first term in the above equation represents the FI contributed by the transparency of $\mathcal{F}^{\Lambda}_{q,res}$. By using the above equation to solve $\partial\mathcal{F}_q/\partial\delta_{ge}=0$, we find the optimal detuning near the resonance center of
\begin{equation}
\begin{aligned}
\delta_{q,opt}^{\Lambda,0}&=0,\\
\delta_{q,opt}^{\Lambda,\pm}&=\pm\frac{1}{2}\sqrt{\frac{3(6+\alpha)\Gamma^2-36|\Omega_c|^2}{[72+\alpha(24+\alpha)]\frac{\Gamma^4}{|\Omega_c|^4}-48(6+\alpha)\frac{\Gamma^2}{|\Omega_c|^2}+228}},
\label{d_ge_opt}
\end{aligned}
\end{equation}
where $\delta_{q,opt}^{\Lambda,0}$ represents the trace of $\mathcal{F}_q^{res}$; and the side-band of $\delta_{q,opt}^{\Lambda,\pm}$ denotes the trace of FI contributed by the absorption. Since demarcation happens when $\mathcal{F}_q^{res}$ dominates the FI spectrum, this means that the condition of $\delta_{q,opt}^{\Lambda,\pm}=\delta_{q,opt}^{\Lambda,0}$, which can give the relation 
\begin{equation}
\alpha_d=12\frac{\left|\Omega_c\right|^2}{\Gamma^2}-6,
\label{alpha_d}
\end{equation}
where $|\Omega_c|^2$ must be greater than $\Gamma/\sqrt{2}$. Eq.\ref{alpha_d} shows that the demarcation point is only dependent on the coupling field strength $|\Omega_c|^2$. The demarcation point of $\alpha_d$ is defined for a chosen coupling field strength $|\Omega_c|^2$ in a $\Lambda$-type system. Furthermore, Eq.\ref{alpha_d} also can be seen as a criterion for distinguishing the region of absorption-dominance and transparency-dominance in the spectral-line-shape. To check the correctness of Eq.\ref{alpha_d}, we compare the result with the numerical calculation. As shown in FIG.\ref{Fig13}, there is a good fit between the results.

The above discussion gives a standard for measuring frequency using the EIT medium. For example, under a fixed coupling intensity, assuming that we need to estimate the frequency away from the resonance, we must keep optical depth, under the condition of less than $\alpha_d$, even selecting a specific optical depth to enhance the information of the frequency estimation we are interested in. If we need to understand the frequency of the near-resonance-center, the medium must be prepared at a condition greater than $\alpha_d$ to ensure that we can obtain the higher FI near the resonance center. The upper limit of the superiority over the quantum method for the frequency estimation of a three-level system will be discussed in the next section.

\subsubsection{Limitation of the quantum-enhanced factor}

\begin{figure*}[!htbp]
\centering
\includegraphics[width=0.9\textwidth]{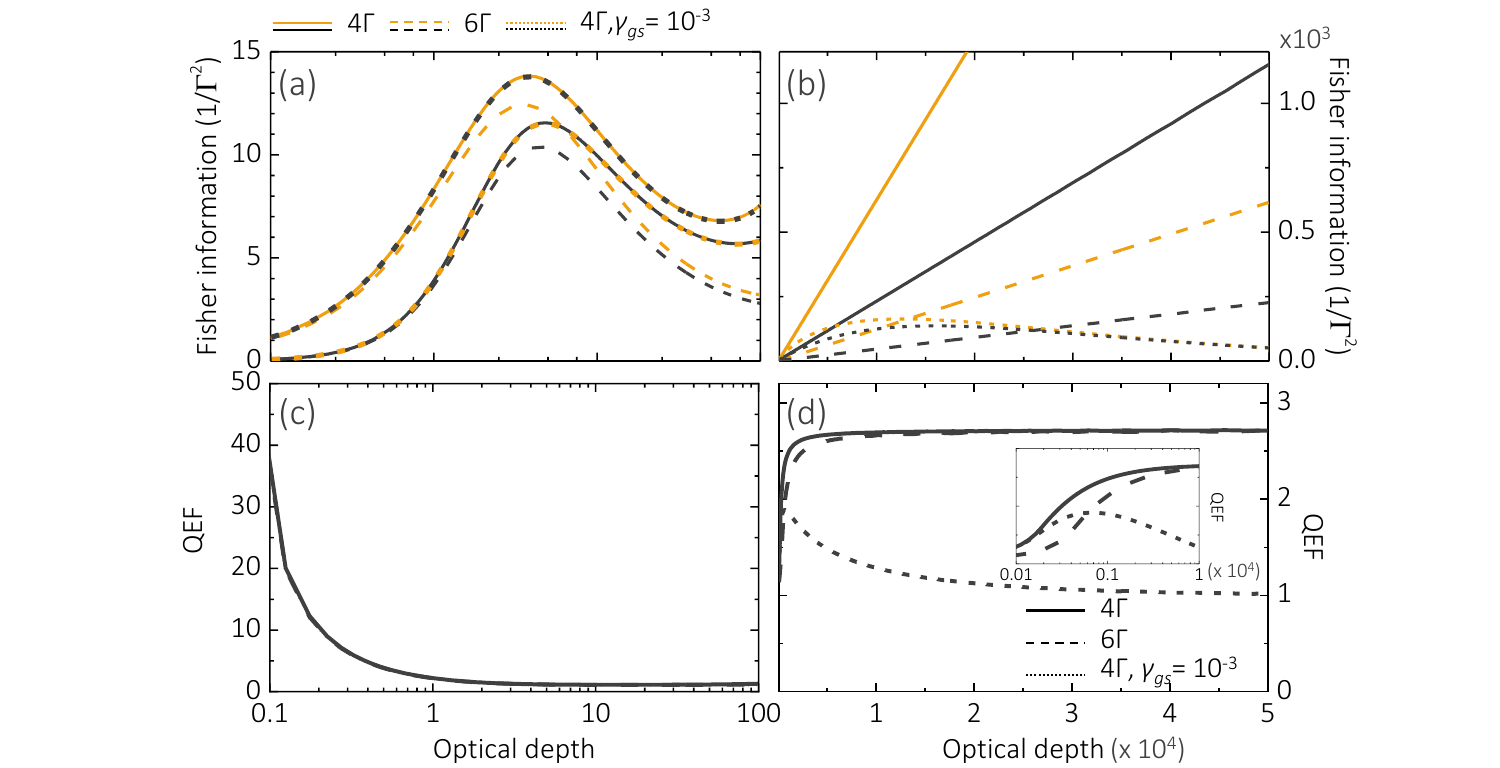}
\caption{The FI and QEF of both cases of quantum (yellow) and classical (black) when detuning allows us to choose an optimal condition at a fixed optical depth. The solid (dashed) lines represent the case when $\Omega_c=4(6)\Gamma$  and without the decoherence rate. The dotted line denotes the cases when $\Omega_c=4(6)\Gamma$ and $\gamma_{gs}=10^{-3}\Gamma$. (a) and (c) show the case of a low optical depth range (sharing the same x-axis), in which the FI is dominated by ATS absorption. Notice that $\gamma_{gs}=0\Gamma$ and $\gamma_{gs}=10^{-3}\Gamma$ almost overlap. (b) and (d) show the high range of the optical depth (sharing the same x-axis), in which the FI is dominated by EIT-transparency.}
\label{Fig14}
\end{figure*}

To compare the QEF at the optimal conditions in the $\Lambda$-type system, we follow the analytical process for the two-level case, i.e. find the optimal FI at the optimal $\delta_{ge}$ for both the quantum and classical cases, then calculate the QEF. To do so, we extract FI with the tracing detuning as shown in FIG.\ref{Fig12} for both cases and further get the QEF. FIG.\ref{Fig14} (a) and (b) show the FI for both cases in the system with different coupling strengths at different optical depth ranges. For the case shown in FIG.\ref{Fig14}, the range of the small $\alpha$ is plotted. In the low $\alpha$ range, it can be seen that the behavior is quite similar to that shown in FIG.\ref{Fig8} in the case of $\Omega_c=6\Gamma$. This phenomenon can be understood as occurring since the structure of the spectrum is dominated by ATS absorption in this region, thus we can get the behaviors similar to those illustrated in FIG.\ref{Fig8}. However, when we look at the maximum value of FI in FIG.\ref{Fig14}, it seems to be about 4 times the maximum value of FI in FIG.\ref{Fig8}. To explain this difference, consider Eq.\ref{Fqopt_a} and Eq.\ref{Fcopt_a}. Since the ATS absorption line shape has a bandwidth half that of two-level absorption \cite{cohen1998atom}, we instinctively modify the bandwidth $\Gamma$ to $\Gamma/2$ in Eq.\ref{Fqopt_a} and Eq.\ref{Fcopt_a}, to obtain the picture the FIG.\ref{Fig14} (a). After modification, the maximum FI for both cases can be easily derived for $\mathcal{F}^{ATS}_{q,max}\approx4\times2.6/\Gamma^{2}$ and $\mathcal{F}^{ATS}_{c,max}\approx4\times2.16/\Gamma^{2}$, which is in good agreement with FIG.\ref{Fig14} (a). For the QEF, we see a similarity in behavior between FIG.\ref{Fig8} and FIG.\ref{Fig14} (c) also, which gives us more confidence in this model.

Notice that in the case of $\Omega_c=4\Gamma$ in FIG.\ref{Fig14} (a), there is an inflection point around $\alpha\sim75$. This behavior indicates the gradual entrance of the system into the transparency-dominated region. According to Eq.\ref{alpha_d}, the demarcation point  $\alpha_d$ is 186 in the $\Omega_c=4\Gamma$ case. Observe the inflection point near the end of the x-axis in  FIG.\ref{Fig14}. Following the increase in the optical depth, the $\Lambda$-type system gradually becomes transparency-dominated, which means that, in quantum cases, the FI will be dominated by $\mathcal{F}^{\Lambda}_{q,res}$. This explains why the trend of the quantum FI in FIG.\ref{Fig14} is linear. It is an interesting result that, in the classical case, as shown in FIG.\ref{Fig14} (b), it also follows a linear trend. Even though the linear trend means that FI can approach infinity, which means an extremely precise parameter estimation in both the quantum and classical cases. However, the trend also suggests that there will be a limitation to the QEF in the high-optical-depth region. FIG.\ref{Fig14} (d) shows the QEF in large $\alpha$ cases. We can clearly see that the QEF is limited to a fixed value for all $\Omega_c$ in the high $\alpha$ region.

In order to find the limitation, we first estimate the FI in the high  $\alpha$ region. Based on the above, the FI in the quantum case is dominated by $\mathcal{F}^{\Lambda}_{q,res}$, hence we only need to calculate FI within the high $\alpha$ region for the classical method. FIG.\ref{Fig11} and FIG.\ref{Fig12} show that the optimal FI in the classical case approaches $\delta_{ge}=0$. We first calculate $\partial\mathcal{F}_c/\partial\delta_{ge}$ then expand that at  $\delta_{ge}=0$, then finally solve the equation for $\lim_{\alpha\rightarrow0}\partial\mathcal{F}_c/\partial\delta_{ge}=0$ to arrive at the solution of

\begin{equation}
\begin{aligned}
\delta^{\Lambda,0}_{c,opt}&=0,\\
\delta^{\Lambda,\pm}_{c,opt}&=\pm\frac{1}{2}\frac{|\Omega_c|^2/\Gamma}{\sqrt{\alpha-\alpha_d}},
\label{d_ge_copt}
\end{aligned}
\end{equation}
where $\alpha>\alpha_d$. By sub $\delta^{\Lambda,\pm}_{c,opt}$ into $\mathcal{F}_c$ and consider $\alpha\gg\alpha_d$, we obtain the FI in the classical case at the high $\alpha$ region of
\begin{equation}
\begin{aligned}
\lim_{\alpha\rightarrow\infty}\mathcal{F}_c(\delta^{\Lambda,\pm}_{c,opt})=\frac{16\alpha\Gamma^2}{e|\Omega_c|^4}. 
\label{Fc_lim_a}
\end{aligned}
\end{equation}
Eq.\ref{d_ge_copt} shows that following the increase in optical depth, the classical method provides a more precise frequency estimation for the near-resonance-center of the spectrum of $\Delta\omega_{EIT}/2\sqrt{ln2}$. Notice that, however, the classical method cannot always give the information at the absolute resonance center of the spectrum. In terms of the limitation of QEF, compare Eq.\ref{FI_lim} and Eq.\ref{Fc_lim_a}, to obtain a straight-forward result indicating that the limitation of QEF is Euler’s number $e\approx2.71$.

To illustrate the advantage of the quantum method, we further compare the QEF, but with the optimal conditions chosen with the classical method.	By substituting Eq.\ref{d_ge_copt} into Eq.\ref{Fqd} and considering $\alpha\gg\alpha_d$, we arrive at $\mathcal{F}_{q}(\delta^{\Lambda,\pm}_{c,opt})=16\alpha\Gamma^{2}/(e-1)\left|\Omega_c\right|^4$. Therefore, even with the optimal conditions chosen for the classical case, the quantum method still provides a QEF of $e/(e-1)\approx1.6$, which shows the limitation of the quantum advantage for estimating the frequency of the near-resonance-center.

\begin{figure}[t]
\centering
\includegraphics[width=0.43\textwidth]{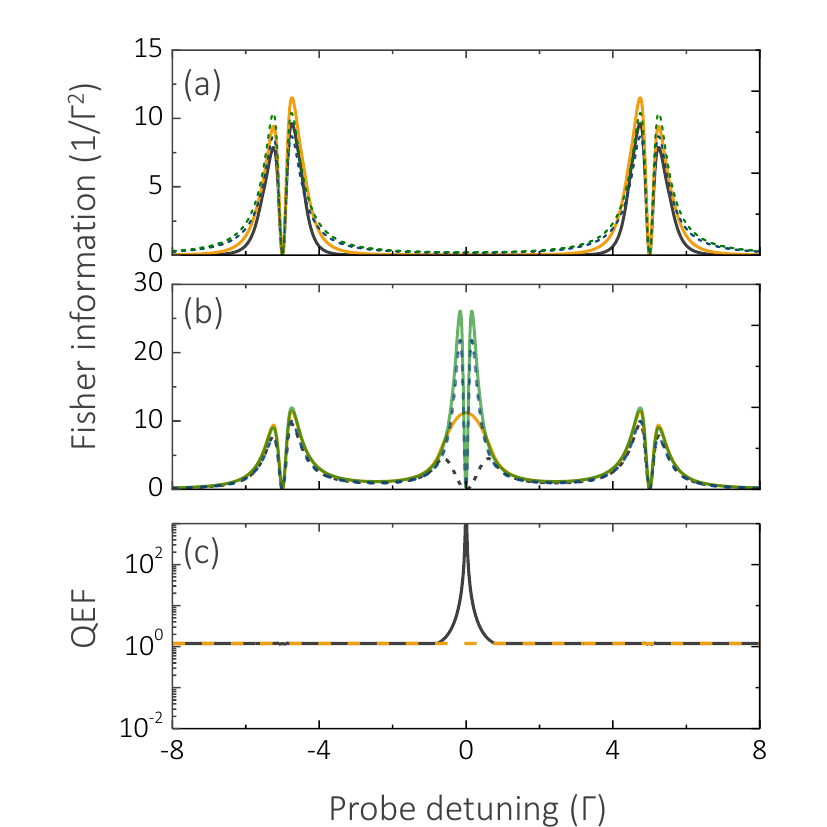}
\caption{FI in the quantum (yellow) and classical (black) cases while adjusting the optical depth for a $\Lambda$-type system. The coupling field is set at $\Omega_c=10\Gamma$. (a) The optical depth range is set from 0 to 10. The green (blue) dotted line represents the absorption FI in quantum (classical) fitting. (b) The optical depth range is set from 0 to 7000. The yellow solid line and black dashed line represent the FI in the quantum and classical cases, respectively, given the condition $\gamma_{gs}=0$. The green solid line and blue dashed line denote the FI in the quantum and classical cases, respectively, when the condition is $\gamma_{gs}=10^{-3}\Gamma$. (c) The quantum-enhanced factor of (b). The black solid line represents the case of $\gamma_{gs}=0$. The yellow dash line represents the case of $\gamma_{gs}=10^{-3}\Gamma$  and the curve is a constant of $\approx1.2$.}
\label{Fig15}
\end{figure}

Now consider that the system allows the optical depth to be adjusted in order to find an optimal FI at a fixed $\delta_{ge}$. Unlike the case with the two-level system, in the $\Lambda$-type system there are no limitations around the resonance canter as $\alpha$ increases (see Eq.\ref{FI_lim} and Eq.\ref{Fc_lim_a}). Therefore, this should give a bound to the optical depth for finding the optimal FI in this case. To show this $\alpha$-dependent behavior, we plot two different finite ranges of $\alpha$ to find the optimal FI, as shown in FIG.\ref{Fig15}. We can see the FI after adjusting the optical depth. There is always a contribution by the ATS absorption at around $\delta_{ge}=\pm\Omega_c/2$. Once again, we can model the behavior by applying Eq.\ref{Fqopt_a} and Eq.\ref{Fcopt_a} but the bandwidth and position will be changed to $\Gamma/2$ and $\delta_{ge}=\pm\Omega_c/2$, respectively. As shown by the  green and blue dotted lines in FIG.\ref{Fig15} (a), there is good agreement with the numerical results. Furthermore, since the range of optical depth is too small to support an $F_q^{res}$ greater than $\mathcal{F}^{ATS}_{q,max}=4\times2.6/\Gamma^2$, the dominance of $\mathcal{F}^{ATS}_{q,max}$ in FIG.\ref{Fig15} (a) still holds. To break through the limitation of absorption-dominance of FI (i.e. $\mathcal{F}^{\Lambda}_{q,res}\geq4\times2.6/\Gamma^2$), we have to apply a greater optical depth of $0.65\Omega_c^4/\Gamma^4$. In FIG.\ref{Fig15} (b), we demonstrate the case of $\Omega_c=10\Gamma$ that corresponds to the critical optical depth of 6500. It can therefore be seen that the FI spectrum is dominated by $\mathcal{F}^{res}_{q}$.

An interesting result is given for the QEF when the system allows the optical depth to be adjusted. In the $\Lambda$-type system, the FI is contributed by the two parts of transparency and absorption. For the absorption part, it has been proven that there is a limitation of 1.2 (Eq.\ref{Q_opt}). This result can also be observed in FIG.\ref{Fig15} (c) at a detuning far from resonance. For the transparency component, however, the QEF shows a clear advantage approach to infinity. Such results are given, on the one hand, because the classical estimation method cannot analyze the information at the resonance center, and on the other hand, while the FI in the quantum method is not limited as optical depth increases at the resonance center. The key is in the nature of the latter, that the transmission of the transparency window is always 100\% regardless of optical depth in the case of $\gamma_{gs}=0$. The optical depth only affects the bandwidth of the transparency window and therefore gives a result with no limitation. In a real system, the existence of a finite decoherence rate is possible, which will be discussed in the next section.

\subsubsection{Finite decoherence rate}

When the system has a finite decoherence rate $\gamma_{gs}$, this makes the transmission of transparent windows become imperfect in the $\Lambda$-type system. In this section, we discuss how an imperfect $\Lambda$-type system impacts the behavior described in the sections above.

\begin{figure}[t]
\centering
\includegraphics[width=0.48\textwidth]{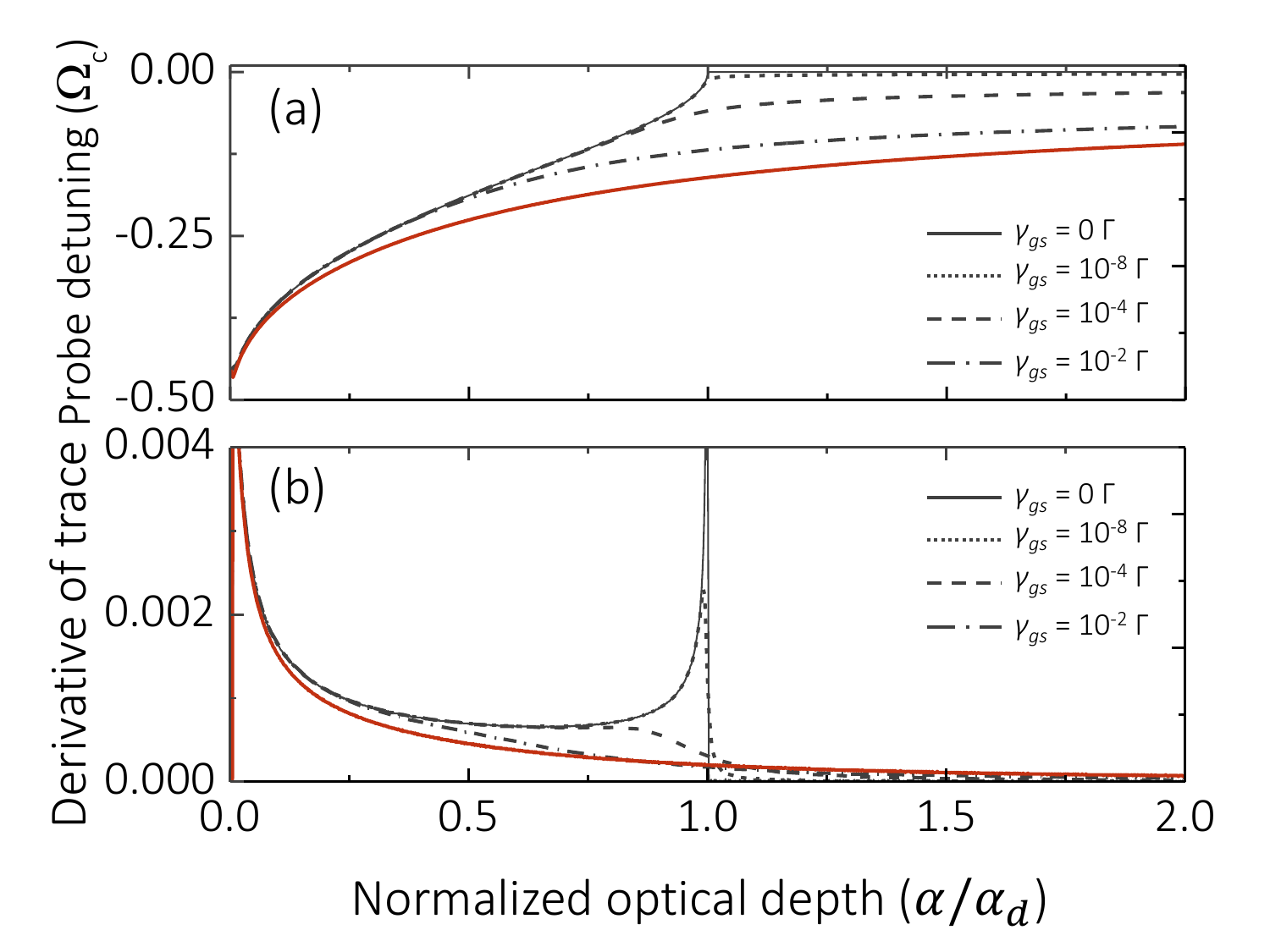}
\caption{(a) Tracing of the maximum Fisher information in the quantum (black) and classical (red) cases with a series of different $\gamma_{gs}$. (b) Derivative behaviors of the FI-trace. In the classical case, both the traces and derivative almost overlap for all $\gamma_{gs}$.}
\label{Fig16}
\end{figure}

The first impact is the disappearance of demarcation $\alpha_d$ in the trace of the $\mathcal{F}_{q}(\delta_{ge})$ diagram. Consider a series of different $\gamma_{gs}$ then trace $\mathcal{F}_{q}(\delta_{ge})$, as shown in FIG.\ref{Fig16} (a). We can see there is no clear demarcation present in the trace of $\mathcal{F}_{q}(\delta_{ge})$ at an optical depth of $\alpha_d$. However, there is still a clear inflection present in the trace of $\mathcal{F}_{q}$ around $\alpha_d$. This means that there are two sources of contribution to  $\mathcal{F}_{q}$, with different behaviors (i.e. absorption and transparency) in the case of a finite $\gamma_{gs}$.

To analyze the relationship between $\gamma_{gs}$ and the inflection, consider the derivative behaviors shown in the FI-trace of $\partial\delta^{\Lambda,-}_{q,opt}/\partial\alpha$ in FIG.\ref{Fig16} (b). We can see that $\partial\delta^{\Lambda,-}_{q,opt}/\partial\alpha$ shows a clear transition between the region of dominated by absorption and transparency. The numerical results obtained by finding the local maximum of $\partial\delta^{\Lambda,-}_{q,opt}/\partial\alpha$ around $\alpha_d$, show that after  $\gamma_{gs}\geq10^{-3}\Gamma\equiv\gamma_{gs,c}$, there are no local maximums, which means that the transition behavior has been eliminated when $\gamma_{gs}\geq\gamma_{gs,c}$. In the classical method, the FI is contributed by absorption, therefore, the structure of the traces of FI in classical cases are almost identical, as shown FIG.\ref{Fig16}.

Interestingly, $\gamma_{gs,c}$ is a universal value for the FI-traces. This result can be seen in FIG.\ref{Fig16} (a),  where we have already considered the nondimensionalization of the optical depth and $\delta_{ge}$ with a unit of $\alpha_d$ and $\Omega_c$ in the FI-trace diagram. In this presentation, the trace of FI is independent of $\Omega_c$. The structure of the trace is determined by a single parameter of $\gamma_{gs}$.

Regarding the impact of $\gamma_{gs}$ on the QEF, a finite $\gamma_{gs}$ mainly affects the transparent window transmission but has no effect on the ATS absorption. Therefore, the absolute FI values are the same in both cases, where $\gamma_{gs}=0$ and $\gamma_{gs}>0$, in the low optical depth region in which the FI is dominated by absorption. This also results in the same QEF values [see FIG.\ref{Fig14} (a) and (c)]. However, since the FI is led by transparency in the high optical depth region, the impact from $\gamma_{gs}$ becomes significant in this region. FIG.\ref{Fig14}2 (b) and (b), illustrates the case for a finite $\gamma_{gs}=10^{-3}\Gamma$. The decay of transmission of the transparent windows [$T(0)=e^{-\frac{\alpha}{1+|\Omega_c|^2/\Gamma\gamma_{gs}}}$], means that there is a decrease in the Fisher information for transmission estimation as the optical depth increases. However, if we consider the spectral line shape of EIT, a higher optical depth corresponds to a narrower transparent window bandwidth, which  is superior for frequency estimation. Taken altogether, the optimal FI will cause a peak in the finite $\gamma_{gs}$ case and its linear behavior will be lost as a function of the optical depth. For the QEF, following the increase in the optical depth, the transmission loss (absorption) dominates the FI meaning that at high optical depths, the FI tends to have the same value in both the quantum and classical cases. Therefore, the QEF also has a local maximum, as shown FIG.\ref{Fig14} (d).

Unlike the case of $\gamma_{gs}=0$, FIG.\ref{Fig14} implies the existence of an optimal value for maximizing FI at each detuning, which is similar to the two-level case. To find these conditions, we solved the optimal $\alpha$ for both the classical and quantum cases, by the equations $\partial\mathcal{F}_q/\partial\alpha=0$ and $\partial\mathcal{F}_c/\partial\alpha=0$. We further found

\begin{widetext}
\begin{equation}
\alpha^\Lambda_{c,opt}=\frac{2\left[(\Gamma^2+4\delta_{ge}^2)(\gamma_{gs}^2+4\delta_{ge}^2)+2(\Gamma\gamma_{gs}-4\delta_{ge}^2)|\Omega_c|^2+|\Omega_c|^4 \right]}{\Gamma\left[\Gamma(\gamma_{gs}^2+4\delta_{ge}^2)+\gamma_{gs}|\Omega_c|^2 \right]},
\label{alpha_copt_lam}
\end{equation}
and 
\begin{equation}
\alpha^\Lambda_{q,opt}=\alpha_{c,opt}^{\Lambda}\left[1+\frac{1}{2}W(-\frac{2\eta}{e^2})\right],
\label{alpha_qopt_lam}
\end{equation}
where $\alpha^\Lambda_{c,opt}$ and $\alpha^\Lambda_{q,opt}$ denote the optimal optical depth for the classical and quantum cases, respectively. Note that $\alpha^\Lambda_{c,opt}$ is independent on $\eta$. Inserting $\alpha^\Lambda_{c,opt}$ and $\alpha^\Lambda_{q,opt}$ into Eq.\ref{Fcd} and Eq.\ref{Fqd} we find the optimal FI in both the classical and quantum cases, as follows:

\begin{equation}
\mathcal{F}_{c,\alpha}^\Lambda=\frac{256\delta_{ge}^2\eta[\Gamma(\gamma_{gs}^2+4\delta_{ge}^2)^2-\gamma_{gs}(\Gamma^2-2\Gamma\gamma_{gs}+\gamma_{gs}^2+8\delta_{ge}^2)|\Omega_c|^2-(\Gamma+2\gamma_{gs})|\Omega_c|^4]^2}{e^2 [\Gamma (\gamma_{gs}^2+4 \delta_{ge}^2) + \gamma_{gs}|\Omega_c|^2]^2 [(\Gamma^2+4\delta_{ge}^2) (\gamma_{gs}^2+4\delta_{ge}^2)+2(\Gamma \gamma_{gs}-4\delta_{ge}^2)|\Omega_c|^2+|\Omega_c|^4]^2},
\label{F_copt_lam}
\end{equation}
\end{widetext}
and 
\begin{equation}
\mathcal{F}_{q,\alpha}^\Lambda=Q^T_{\alpha,opt}\mathcal{F}_{c,\alpha}^\Lambda.
\label{F_qopt_lam}
\end{equation}
As can be seen in Eq.\ref{F_qopt_lam}, since a finite $\gamma_{gs}$ now exists in the system, the QEF is limited by the absorption limitation of $Q_{\alpha,opt}^T\approx1.2$, which is the same as for the two-level case and
independent of detuning. To demonstrate the behavior, we plot the optimal FI for both cases when $\gamma_{gs}=10^{-3}\Gamma$, as shown FIG.\ref{Fig15} (b).

Compared to the case of $\gamma_{gs}=0$ (see FIG.\ref{Fig15} (c)), the difference between the QEFs is due to the system now having the optimal optical depths of $\alpha^\Lambda_{q,opt}$ and $\alpha^\Lambda_{c,opt}$. Therefore, the system does not allow an unlimited increase in the optical depth for enhancing the QEF on $\delta_{ge}=0$. A way to break through the limitation of $Q_{\alpha,opt}^T$ on the resonance center is to modify the strength of the coupling field and further narrow down the bandwidth of the transparency window. A narrower window provides for a more accurate frequency estimation at the resonance center than for the higher coupling case, which provides higher Fisher information and QEF.

\subsubsection{Technical loss}

In the previous section, we considered the impact on frequency estimation of the imperfection of the sample. In this section, we  discuss the case where the measurement method has a finite technical loss but without a decoherence rate. As can be seen in FIG.\ref{Fig15} (c), there is basically no limitation to the QEF  at the resonance center in the perfect case (i.e. $\eta=1$ and $\gamma_{gs}=0$). Note that, in the case when the FI is contributed by absorption, the QEF has a limitation of 1.2 according to Eq.\ref{Q_opt}, therefore, there is a big advantage to using the quantum method in the $\Lambda$-type system. However, this advantage will be significantly less when the technical loss is involved. To illustrate the behavior, consider a QEF at the resonance center of $\delta_{ge}=0$. According to Eq.\ref{Fcd} and Eq.\ref{Fqd}, the QEF is given by
\begin{equation}
Q^{\Lambda}_{\alpha,opt}=\left(\frac{\alpha_{q,opt}^{\Lambda}}{\alpha_{c,opt}^{\Lambda}}\right)^2\frac{1}{1-\eta},
\label{QEF_alpha1}
\end{equation}
where $\alpha_{q,opt}^{\Lambda}$ and $\alpha_{c,opt}^{\Lambda}$ are the optimal optical depths in both cases. For an equitable evaluation of the QEF, we follow the ratio of the optical depth applied in Eq.\ref{alpha_qopt_lam}. Therefore, the QEF with finite technical loss is given by
\begin{equation}
Q^{\Lambda}_{\alpha,opt}=\left[1+\frac{1}{2}W(-\frac{2\eta}{e^2})\right]^2\frac{1}{1-\eta}
\label{QEF_alpha2}
\end{equation}

\begin{figure}[t]
\centering
\includegraphics[width=0.48\textwidth]{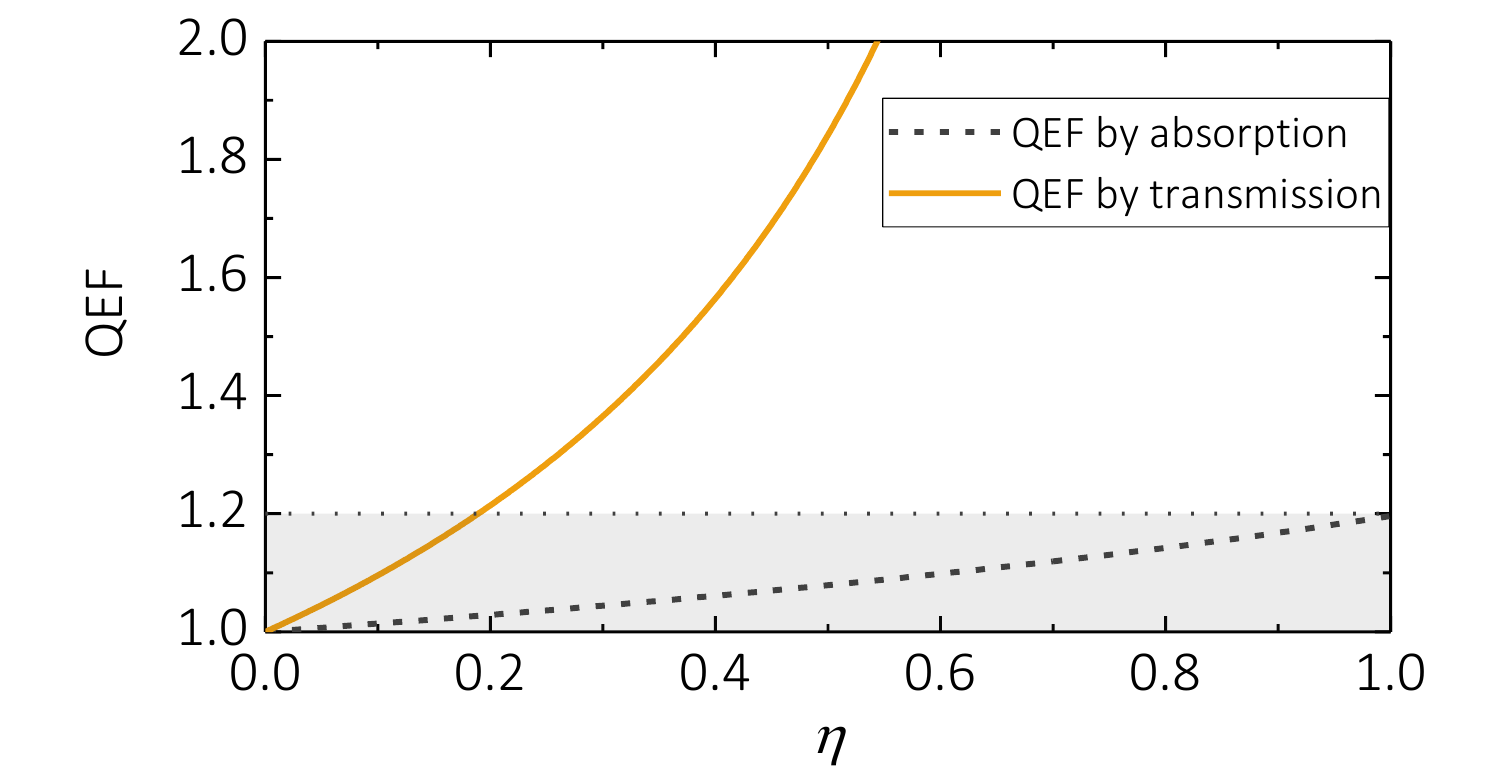}
\caption{Comparison of QEF that is contributed by absorption (black) and transparency (yellow). The gray area represents a limitation $Q_{\alpha,opt}^{T}$ of 1.2. It can be seen that $Q_{\alpha,opt}^{\Lambda}$ breaks through the limitation at $\eta\approx0.187$.
}
\label{Fig17}
\end{figure}

Now we can clearly see that $Q^{\Lambda}_{\alpha,opt}$ can easily surmount $Q^{T}_{\alpha,opt}$ for all $\eta$, as shown in FIG.\ref{Fig17}. Furthermore, $Q^{\Lambda}_{\alpha,opt}$ can break through the limitation to $Q^{T}_{\alpha,opt}$ of 1.2 when the technical loss is greater than $\eta\approx0.187$ in the $\Lambda$-type system. This shows the significant advantage of using the quantum method in a $\Lambda$-type system, or when the spectrum has a unitary transparency window.
\\

\section{Summary}\label{secIV}

In this study, we systemically analyzed the Fisher information for frequency estimation in a two-level and a $\Lambda$-type system. In both cases, we analyzed the structure of the FI spectrum by fixing an optical depth and then tracing the optimal FI in the $\alpha-\delta_{ge}$ diagram. We assume that the system allows us to choose an optimal optical depth and further find a maximum FI for analysis of the limitations of both the quantum and classical methods for frequency estimation. Based on the knowledge gained from the two-level case, we further classify the contribution to the FI from absorption and transparency in the $\Lambda$-type system. Analysis of the trace of the maximum quantum FI showed a clear demarcation point to distinguish between the two regions that are absorption-dominated and transparency-dominated, which provides a way to identify the class of the spectral line shape. For the QEF, we show that the transparency-FI can support an infinite QEF at $\delta_{ge}=0$ when $\gamma_{gs}=0$. Even with the inclusion of technical loss in the system, the QEF will not be limited by the absorption. The QEF is 1.2 when the loss is lower than 0.813. ($\eta=0.187$). In realistic systems, the frequency stabilization and linewidth of the photon source need to be considered. The property of the photon source should be comparable with the atomic medium to ensure the efficient interaction between them \cite{tsai2020quantum,wang2019efficient}. Otherwise, the spectrum measurement might be fuzzy \cite{kim2019effect}, resulting in the loss of Fisher information, which is an additional issue that needs to be considered.

In conclusion, our work provides a systematic study of the single-photon EIT-spectrum based on Fisher information analysis. We show the quantum advantages and limitations of single-photon probe analysis. The analysis of the FI structures deepens our understanding of the characteristics of the $\Lambda$-type media spectrum. We believe this work could have applications in quantum metrology based on the EIT medium, especially in cases using EIT mediums to detect weak signals in the environment.\cite{PhysRevA.62.013808,sun2017cavity,zhang2016high,meyer2021optimal,kuan2016large}.

\bibliographystyle{apsrev4-1} 
%

\end{document}